\documentclass[aps,prb,twocolumn,superscriptaddress,floatfix]{revtex4-2}
\usepackage{amsmath,amssymb,graphicx,bm}
\graphicspath{{figures/}}
\usepackage{color}

\usepackage{tikz}

\newcommand{\sunsetdiag}{%
  \mathchoice
    {\vcenter{\hbox{\tikz[baseline=-0.1ex, line width=0.08em]{\draw (0,0) circle (0.45ex); \draw (-0.45ex,0) -- (0.45ex,0);}}}} 
    {\vcenter{\hbox{\tikz[baseline=-0.1ex, line width=0.08em]{\draw (0,0) circle (0.45ex); \draw (-0.45ex,0) -- (0.45ex,0);}}}} 
    {\vcenter{\hbox{\tikz[baseline=-0.08ex, line width=0.06em]{\draw (0,0) circle (0.32ex); \draw (-0.32ex,0) -- (0.32ex,0);}}}} 
    {\vcenter{\hbox{\tikz[baseline=-0.06ex, line width=0.04em]{\draw (0,0) circle (0.22ex); \draw (-0.22ex,0) -- (0.22ex,0);}}}} 
}

\usepackage{hyperref}
\hypersetup{colorlinks=true, citecolor=blue, linkcolor=blue, urlcolor=blue, breaklinks=true}

\begin{document}

\title{The two-particle-irreducible vertex of the two-dimensional lattice $\phi^4$ model
       across the Ising transition}
\author{Lode~Pollet}
\affiliation{Department of Physics and Arnold Sommerfeld Center for Theoretical Physics (ASC), Ludwig Maximilian University of Munich, 80333 Munich, Germany}
\affiliation{Munich Center for Quantum Science and Technology (MCQST), 80799 Munich, Germany}
\date{\today}

\begin{abstract}
We reconstruct the 2PI vertex $\Gamma(k,p;q)$ from Monte Carlo measurements of the connected two-particle correlator
for the two-dimensional single-component $\phi^4$ lattice field theory and follow it across the Ising transition. Resolving the vertex in the irreducible
representations of the point group $C_{4v}$, we find that the instability is driven by the $A_1$
(ferromagnetic) channel at zero transfer, whose leading eigenvalue of the symmetrized
Bethe--Salpeter kernel approaches unity. Substantial $B_1$ (nematic) and
$B_2$ (diagonal nematic) contributions cooperate with $A_1$ across all system sizes, highlighting that the soft sector is
 multidimensional. In real space, the vertex is short-ranged away
from criticality while it develops a power-law tail at the critical point. In the ordered phase, 
the $q=0$ eigenvalue collapses because the ferromagnetic weight has condensed into the
(one-particle-reducible) order parameter (or collective coordinate for a finite system), although finite-momentum fluctuations persist.
By stripping the crossed-channel ladders, we obtain the fully irreducible vertex, which
is a local contact -- to a very good approximation. Inserted into the parquet and Schwinger--Dyson equations, this contact reproduces the Monte Carlo self-energy with an accuracy better than one-tenth of a percent. This provides a first-principles
benchmark of the dynamical local-vertex approximation (D$\Gamma$A).
Additionally, we demonstrate that in the critical region, the physical solution of the parquet equations behaves as a repulsive fixed point, driven initially by a single order-parameter mode.
\end{abstract}

\maketitle

\section{Introduction}
The single-particle propagator $G$ and self-energy $\Sigma$ are the standard objects of a
correlated field theory~\cite{AGD,FetterWalecka,NegeleOrland}, but they carry only part of the fluctuations: the response of the
system to a change in $G$ is governed by the two-particle-irreducible (2PI) vertex
$\Gamma = \delta\Sigma/\delta G$, the kernel of the Bethe-Salpeter equation~\cite{Salpeter1951}. It is the vertex, not the self-energy, that decides which collective channel becomes unstable and where the correlation
length diverges. For the two-dimensional $\phi^4$ model, which is the paradigmatic realization of the Ising
universality class~\cite{Onsager1944}, the transition is ferromagnetic, and one might expect the vertex to be dominated
by the fully symmetric ($A_1$) channel at zero momentum transfer.  Furthermore, the Green function and the self-energy depend only on a single momentum in a translation-invariant system and reside exclusively in $A_1$. However, as we will show, this expectation is incomplete:
As fluctuations strengthen, the vertex develops structure in the other irreducible representations
of the lattice point group, most notably a nematic $B_1$ component, followed by a diagonal nematic $B_2$ component.

In this work, we reconstruct the 2PI vertex $\Gamma(k,p;q)$, which follows by inversion from Monte Carlo measurements of the connected two-particle correlator. We resolve $\Gamma(k,p;q)$
in the irreducible representations of $C_{4v}$, track the leading Bethe-Salpeter eigenvalue as it
approaches unity at the phase transition (also known as the Thouless point) and characterize the vertex both in momentum and in real space. To the best of our knowledge, such a channel- and transfer-resolved map of the
2PI vertex has not been reported in full for the two-dimensional Ising/$\phi^4$ problem. 
We will then extract the fully irreducible vertex of the parquet by stripping the reducible (crossed-channel ladder) contributions from the 2PI vertex, in the disordered phase up to the critical point.
The fully irreducible vertex plays a central role in such theories as the parquet formalism~\cite{Diatlov1957,DeDominicis1964,Bickers1991}, the $n-$PI theories~\cite{Berges2004,Carrington2004}, the Schwinger-Dyson equations (SDE)~\cite{SDE} and the functional renormalization group (fRG)~\cite{RMP_fRG,Berges2002}.
This will allow us to establish D$\Gamma$A~\cite{Rohringer2018} as a very good approximation in the disordered phase, and to analyze the stability of the parquet equations in the critical region, where the physical solution turns into a repulsive fixed point.

\section{Model and method}
We study the single-component $\phi^4$ theory on the square lattice with Euclidean action
\begin{equation}
  S = -\beta\sum_{\langle ij\rangle}\phi_i\phi_j + m^2\sum_i \phi_i^2 + \lambda\sum_i \phi_i^4,
\end{equation}
at $m^2=0$ and $\lambda=\tfrac12$, sampled with the Brower--Tamayo cluster algorithm~\cite{BrowerTamayo1989}. Monte Carlo does
not give $\Gamma$ directly; what is accumulated is the connected two-particle correlator of the
composite operator $\rho_q(k)=\phi_k^{*}\phi_{k+q}$,
\begin{align}
  \chi(k,p;q) &= \langle \rho_q(k)\,\rho_q(p)^{*}\rangle
              - \langle \rho_q(k)\rangle\langle \rho_q(p)^{*}\rangle .
\end{align}
The vertex is obtained by the Bethe--Salpeter equation (BSE)~\cite{Salpeter1951},
\begin{equation}
  \Gamma  =   \chi_0^{-1} - \chi^{-1},
\end{equation}
where $\chi$ is full susceptibility and the bare susceptibility $\chi_0$ is given by
\begin{equation}
  \chi_0(k,p;q)  =   G(k)G(k+q)\,[\delta_{k,p}+\delta_{p,-k-q}],
\end{equation}
which also carries the exchange term $\delta_{p,-k-q}$ required by the fact that the
field is real ($\phi_{-k}=\phi_k^{*}$). The BSE can be used, for instance, to describe collective excitations (such as plasmons and magnons), bound states (such as excitons and Cooper pairs), and thermodynamic response (such as viscosity and thermal conductivity). The 2PI version of the BSE can be shown to be derivable in second order from a Luttinger-Ward functional, which ensures that macroscopic conservation laws such as charge and energy are conserved and that non-perturbative physics can be studied~\cite{BaymKadanoff1961,Baym1962}. We note in passing that the results presented here stand on their own and do not rely on a Luttinger-Ward functional, although they do have implications for theories built on it.

We monitor the instability using the symmetrized kernel given by
\[
K = \chi_0^{1/2} \Gamma \chi_0^{1/2} = 1 - \chi_0^{1/2} \chi^{-1} \chi_0^{1/2},
\]
where the leading eigenvalue reaches unity at the Thouless point. The function \(\rho_q(k)\) satisfies \(\rho_0(k) = \rho_0(-k)\) at \(q=0\) due to the fact that the field \(\phi\) is real. Similar relations hold for other values of \(q\). Consequently, the susceptibility \(\chi\) is rank-deficient by construction. Throughout this work, we focus on its physical (non-null) subspace, where the condition number remains moderate (see Appendix ~\ref{app:stab}). 
At \(q=0\), the operator \(\rho_0(k) = |\phi_k|^2\) is even under inversion, which means that only the four inversion-even irreducible representations \(A_1, A_2, B_1, B_2\) can appear.

\section{Vertex physics of the Ising transition}
\label{sec:vertexphys}

\begin{figure}
  \includegraphics[width=\columnwidth]{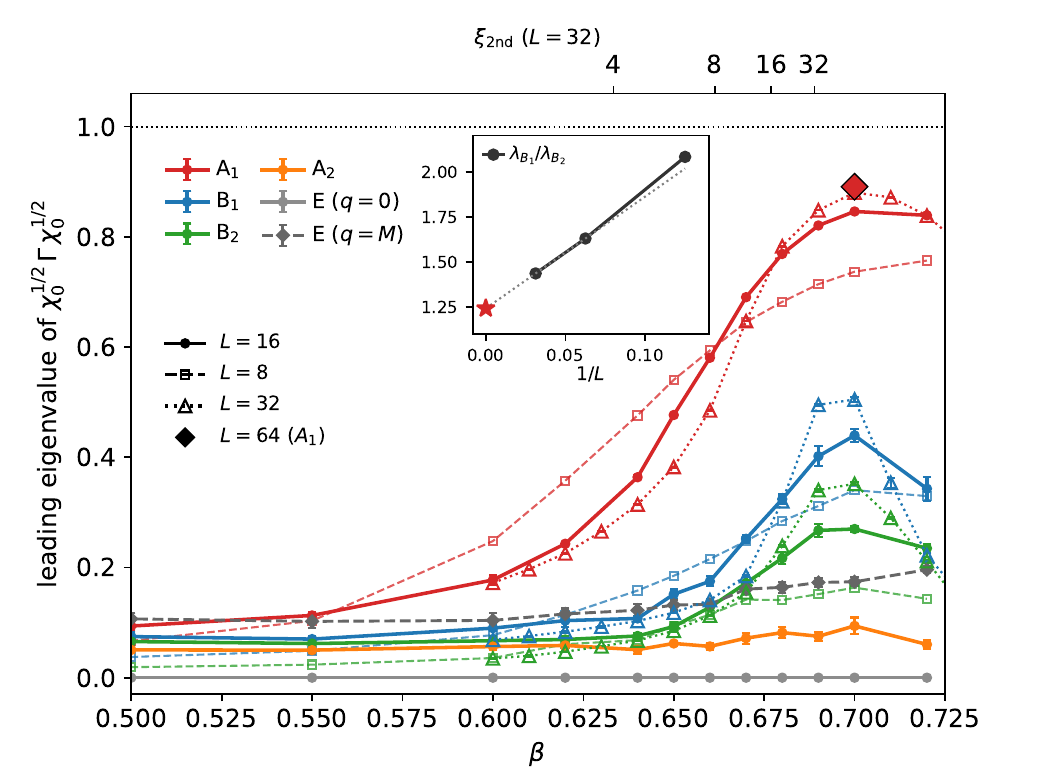}
  \caption{\label{fig:irrep}
  The leading eigenvalue of the symmetrized Bethe-Salpeter kernel, given by $\chi_0^{1/2} \Gamma \chi_0^{1/2}$, is evaluated at zero transfer and resolved in the $C_{4v}$ irreducible representations as a function of the inverse temperature $\beta$ (with $L = 16$, represented by filled symbols). The $A_1$ channel (ferromagnetic) drives the instability toward unity, while the $B_1$ (nematic) and $B_2$ (diagonal nematic) channels cooperate. The $A_2$ channel is inert, and the $E$ channel, associated with anti-ferromagnetic fluctuations, vanishes identically at $q = 0$ due to parity. However, it is non-zero at the corner transfer $q = M = (\pi, \pi)$, indicated by a grey dashed line.
Open symbols illustrate the finite-size trend for the two leading channels: $L = 8$ (dashed) and $L = 32$ (dotted). The $A_1$, $B_1$, and $B_2$ channels all strengthen as $L$ increases, and the peak sharpens as it approaches the critical point. The filled diamond represents the $A_1$ peak for $L = 64$. The upper x-axis shows the second-moment correlation length for comparison (refer to App.~\ref{app:fss}).
In the inset, the ratio $\lambda_{B_1}/\lambda_{B_2}$ at $\beta = 0.70$ is plotted against $1/L$. This ratio decreases toward a finite limit as the two components of the stress-energy tensor align, reflecting the emergent rotational symmetry of the conformal field theory (CFT). The star marks a linear extrapolation of the data in terms of $1/L$.  }
\end{figure}

Figure~\ref{fig:irrep} presents the leading eigenvalues for each symmetry sector. In the disordered phase, all channels are weak and comparable. As we approach the transition, the fully symmetric \(A_1\) channel begins to separate, with its leading eigenvalue increasing toward the Thouless value of unity. This indicates a ferromagnetic instability as observed through the 2PI vertex framework. Initially, the nematic \(B_1\) channel grows alongside \(A_1\), reaching just over half its value at the largest system size.
The finite-size trends for the two leading channels are as follows: for \(A_1\), the values are logarithmically diverging as  \(0.75, 0.85, 0.88, 0.89\) at \(L = 8, 16, 32, 64\); for \(B_1\), the values are \(0.34, 0.44, 0.51\) at \(L = 8, 16, 32\). Note that for \(L = 64\) we could only resolve the leading \(A_1\) eigenvalue because of increasingly demanding statistical requirements.
The \(B_2\) channel behaves similarly to \(B_1\) but exhibits a smaller amplitude, while \(A_2\) remains negligible. Therefore, the soft sector is not characterized by a single mode; instead, it is a cooperating multiplet dominated by \(A_1\). Although only \(A_1\) ultimately diverges, \(B_1\) and \(B_2\) contribute through their off-diagonal coupling with \(A_1\), and should therefore not be omitted. The leading eigenvalues peak slightly beyond the thermodynamic transition point before receding in the ordered phase: the maximal \(A_1\) eigenvalue is observed near \(\beta \simeq 0.70\), while the clean second-moment correlation-length crossing indicates \(\beta_c \simeq 0.685\) (App.~\ref{app:fss}). Therefore, the vertex signal slightly overestimates \(\beta_c\) for finite system sizes; we adopt \(\beta_c \simeq 0.685\) throughout our analysis. It is worth mentioning that stripe and even nematic phases are predicted for the Ising model, but they require additional terms, such as a sufficiently strong antiferromagnetic coupling (which leads to frustration) along the diagonal and the presence of an external magnetic field~\cite{Guerrero2015}.

The $E$ channel is exactly zero at $q=0$ due to parity. However, it takes on a finite value at the corner transfer $M=(\pi,\pi)$, which describes a short-range antiferromagnetic fluctuation. This fluctuation increases gradually through the critical window without nearing instability.

\begin{figure}
  \includegraphics[width=\columnwidth]{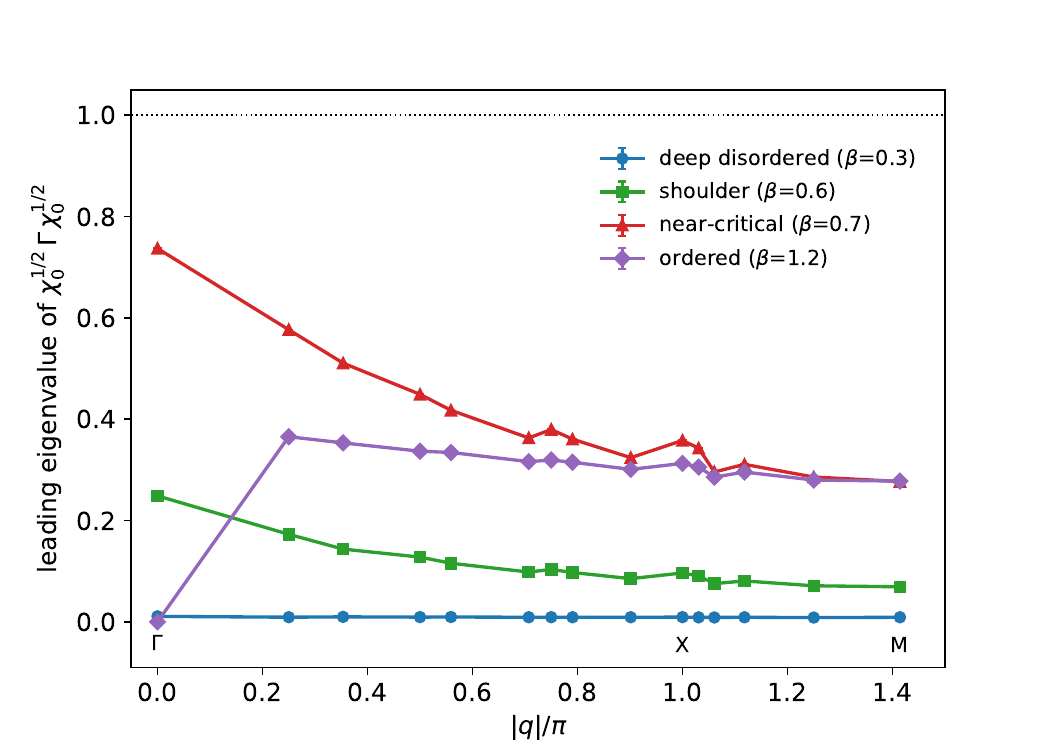}
  \caption{\label{fig:qdep}
  The leading kernel eigenvalue as a function of transfer momentum \( |q| \) is shown along the irreducible wedge, with points marked at \( \Gamma \), \( X \), and \( M \), for four regimes at \( L = 8 \) in the \( A_1 \) sector. The near-critical curve is taken at \( \beta = 0.7 \), where the \( A_1 \) eigenvalue reaches its maximum on this finite lattice of size \( L = 8 \). This finite-size pseudo-critical point is slightly above the thermodynamic critical point \( \beta_c \approx 0.685 \) (as discussed in Sec.~\ref{sec:vertexphys} and App.~\ref{app:fss}). It is important to note that we are plotting this finite-size maximum instead of \( \beta_c \).  In the deep disordered phase, the vertex is weak and nearly independent of \( q \). As we approach criticality, the dominance of the \( q = 0 \) mode increases, while the decay with respect to \( |q| \) becomes more gradual. A band of small-\( q \) modes softens collectively as the correlation length diverges. In the ordered phase (\( \beta = 1.2 \)), the eigenvalue at \( q = 0 \) collapses to zero, while finite-\( q \) fluctuations continue to persist.}
\end{figure}

Figure~\ref{fig:qdep} illustrates the behavior of the vertex as a function of momentum transfer. In the disordered phase, the leading eigenvalue is small and relatively flat across different values of \(|q|\), indicating that the vertex is weak and lacks structure. As we approach criticality, a prominent maximum at \(q=0\) emerges, but the decline away from this point is gradual. This results in a band of small-momentum modes becoming softer, which serves as the momentum-space signature of the diverging correlation length.

In the ordered phase, the scenario at the origin reverses: at \(\beta=1.2\), the eigenvalue at \(q=0\) drops to zero. The ferromagnetic weight that facilitated the transition has condensed into the uniform order parameter (or a collective coordinate for a finite system), which is one-particle-reducible and consequently absent from the 2PI vertex. However, the loss of weight is not limited to the single point \(q=0\); rather, the entire small-\(q\) region becomes depleted. At \(\beta=1.2\), the eigenvalue in the ordered phase is reduced to approximately two-thirds of its critical value at the smallest momentum resolved on the \(L=8\) lattice (\(|q|=\pi/4\)), while the curves converge near the zone boundary.

For the larger \(L=16\) lattice, we are able to access smaller values of \(|q|=\pi/8\), where the suppression intensifies, approaching one-half (data not shown). Thus, the short-wavelength structure of the vertex remains largely unchanged as we enter the ordered phase, whereas the long-wavelength component is suppressed. The gradual depletion of weight as \(|q| \to 0\) is physically expected, as ordering suppresses long-wavelength fluctuations while preserving short-wavelength fluctuations.


The comparison with the disordered side is quite striking. For $\beta = 1.2$, the absolute value of the relative detuning from $\beta_c$, calculated as \(\left|\beta - \beta_c\right|/\beta_c \approx 0.71\), is significantly greater than that for \(\beta = 0.6\), where \(\left|\beta - \beta_c\right|/\beta_c \approx 0.14\). However, the finite-$q$ eigenvalues for \(\beta = 1.2\) are consistently larger by about a factor of two at the smallest momenta, measuring \(0.37\) compared to \(0.17\) for \(L = 8\).

This indicates that the finite-$q$ fluctuation structure gaps out remarkably slowly on the ordered side. While the order parameter condenses sharply in the \(q=0\) channel, the short-wavelength fluctuations are only slightly reduced and retain most of their critical strength well into the ordered phase. In contrast, these fluctuations have not yet developed on the disordered side.

This observation does not contradict the fact that the connected correlation length \(\xi_c\) on the ordered side, as shown in Figure \ref{fig:fss} and Appendix \ref{app:fss}, decreases much more rapidly as a function of detuning compared to the disordered side. For instance, \(\xi_c \approx 2.3\) at the shoulder (\(\beta = 0.6\)) drops to only \(\xi_c \approx 0.4\) at \(\beta = 1.2\). 

In the decomposition \(K = \chi_0^{1/2} \Gamma \chi_0^{1/2}\), the vertex \(\Gamma(q)\) remains essentially flat across all temperatures, while its overall magnitude increases significantly—by roughly a factor of fifty—from the shoulder into the ordered phase. Therefore, all the transfer dependence of \(K\) is contained within the bubble \(\chi_0(k;q) = G(k) G(k+q)\), whose profile is determined by the connected correlation length \(\xi_c\): for long \(\xi_c\), it is strongly \(q\)-dependent (enhanced at small transfer), while for short \(\xi_c\), it is weakly \(q\)-dependent.

\begin{figure}
  \includegraphics[width=\columnwidth]{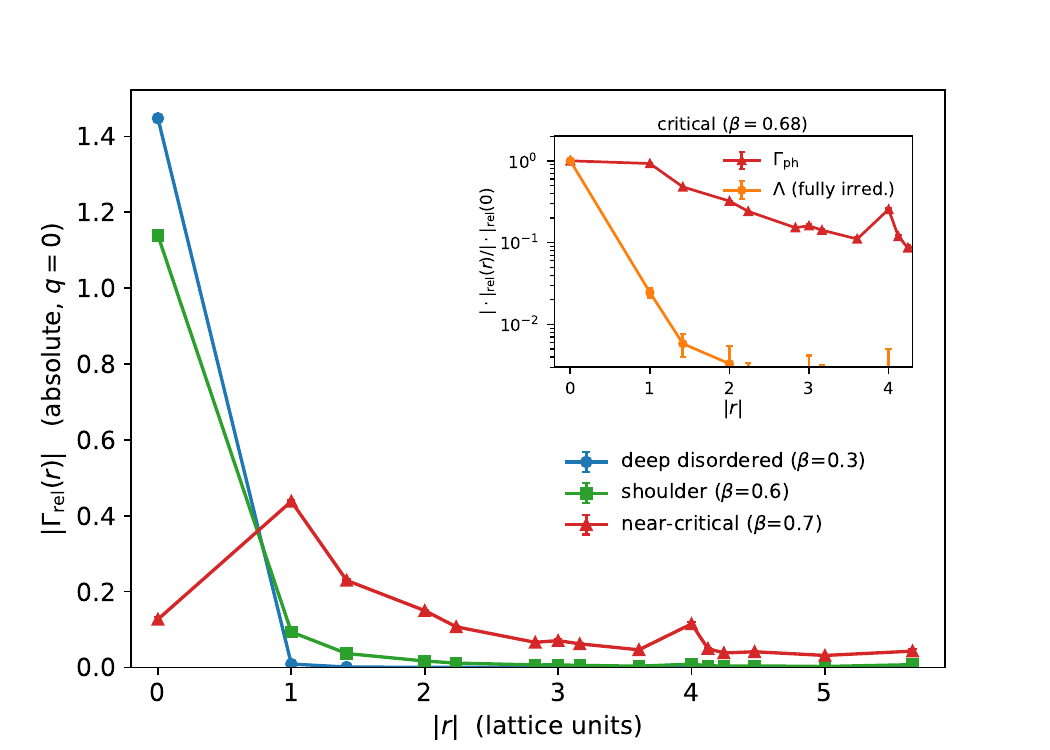}
  \caption{\label{fig:loc}
  The absolute value of the relative-coordinate vertex kernel is defined as $ \Gamma_{\mathrm{rel}}(r) = \frac{1}{N}\sum_{k,p} e^{-i(k-p)\cdot r} \Gamma(k,p;0) $ as a function of lattice distance \( |r| \) for three regimes (with \( L=8 \) on a linear scale). As shown in Fig.~\ref{fig:qdep}, the near-critical curve occurs at \( \beta = 0.7 \), which corresponds to the maximum finite-size \( A_1 \)-eigenvalue for the \( L=8 \) lattice, slightly above the thermodynamic critical point \( \beta_c \simeq 0.685 \). When moving away from criticality, the vertex exhibits a sharp peak at the origin and has negligible weight for \( |r| > 1 \). At criticality, the maximum weight shifts away from the origin to \( |r| = 1 \), and a weak tail begins to develop at larger values of \( |r| \) (as quantified in the text). The enhancement observed at \( |r| = L/2 \) is not limited to \( |r| = 4 \) but is rather a finite-size \( X = (\pi,0) \) zone-boundary effect. The inset shows that at criticality (\( \beta = 0.68 \) for \( L=8 \)), the fully irreducible vertex \( \Lambda \) (described in Sec.~\ref{sec:parquet}) takes on a pure contact form, with no tail distinguishable from the noise. Error bars in both panels are estimated using bootstrap methods over the Monte Carlo vertex bins. Additionally, the role of the center-of-mass coordinate is explained in App.~\ref{app:caveat}.}  
\end{figure}

In Figure~\ref{fig:loc}, we analyze the locality of the 2PI vertex in real space (refer to App.~\ref{app:caveat} for a discussion on the center-of-mass coordinate). The $x$-axis represents the relative coordinate conjugate to the momentum difference \(k - p\), evaluated at zero transfer (\(q = 0\)) and summed over the center-of-mass momentum \(k + p\).

When away from criticality, the kernel shows a sharp on-site spike, which has effectively decayed by one lattice spacing. This indicates that the vertex is ultra-local, primarily influenced by its \(r = 0\) component, consistent with the nearly structureless momentum dependence depicted in Fig.~\ref{fig:qdep}. As we approach the transition, the weight begins to redistribute: the on-site component decreases, and the maximum shifts outward to \(|r| = 1\), with a faint tail extending to larger distances.

High-statistics data at \( L=16 \) indicate that the tail is best described by a power law of the form \( \sim r^{-a} \). The effective exponent softens as it approaches the transition, with values of \( a \simeq 2.4, 2.0, 1.8 \) at \( \beta = 0.64, 0.66, 0.68 \), respectively. This decay is significantly faster than the power-law decay of the propagator, which follows \( r^{-\eta} \) with \( \eta = 1/4 \) at the critical point.

\section{Emergent symmetries}

The three cooperating channels illustrated in Fig.~\ref{fig:irrep} are not an arbitrary selection. For those familiar with conformal field theory, these channels are the lattice analogs of the low-lying even (spin-neutral) operators of the Ising CFT: the energy density ($A_1$) and the two components of the stress-energy tensor ($B_1$, $B_2$). For readers acquainted with diagrammatic expansions, the channel $B_1$ is initially produced in the 2PI vertex by amputating the sunset diagram, while $B_2$ is first generated in higher order.

\subsection{Single-channel identifications}
The fully symmetric \( A_1 \) channel is built on the harmonic \( \cos k_x + \cos k_y \), i.e.\ the isotropic nearest-neighbor energy density \( \varepsilon \), and is therefore explicitly conjugate to the inverse temperature \( \beta \). Its response is the energy (specific-heat) susceptibility, which in the two-dimensional Ising universality class diverges logarithmically at criticality, as known exactly from Onsager's solution~\cite{Onsager1944}. Consistently, \( A_1 \) is the channel that grows and becomes size-dependent on approach to the transition.

The nematic \( B_1 \) channel is built on \( \cos k_x - \cos k_y \), conjugate to the difference of the two nearest-neighbor couplings \( J_x - J_y \), i.e.\ the lattice-anisotropy direction. The corresponding response is the anisotropy susceptibility, given in the Ising class by the second derivative of the free energy with respect to the anisotropy at the isotropic point (located by Onsager's anisotropic solution \( \sinh(2\beta J_x)\sinh(2\beta J_y)=1 \)). The anisotropic susceptibility will remain finite at the transition. Note that the Ising universality class guarantees us a logarithmically diverging $A_1$ and finite $B_1$ response for $\phi^4$, but the values will not match identically with Onsager's solution for the 2D Ising model.


\subsection{Emergent rotational symmetry}
The $B_2$ channel, built on $\sin k_x \sin k_y$, has no Onsager counterpart. At small momentum $\cos k_x - \cos k_y \sim k_x^2 - k_y^2$ and $\sin k_x \sin k_y \sim k_x k_y$ are the real and imaginary parts of $(k_x + ik_y)^2$: $B_1$ and $B_2$ are the two components of the traceless stress-energy tensor, anisotropic strain $B_1 \sim T_{xx} - T_{yy}$ and shear $B_2 \sim T_{xy}$. On the square lattice, they belong to distinct $C_{4v}$ irreducible representations---no lattice symmetry relates them, since the $45^\circ$ rotation that would, is not in the point group---so, a priori, they fluctuate independently, and $B_2$, having no bare coupling, is generated entirely by fluctuations. Continuous rotational invariance, emergent at the critical point, nonetheless locks the two together into a fixed ratio.

We observe this locking directly: the eigenvalue ratio $\lambda_{B_1}/\lambda_{B_2}$ decreases monotonically with system size toward a finite limit ($2.08, 1.63, 1.44$ at $L=8, 16, 32$; see the inset of Fig.~\ref{fig:irrep}), a linear $1/L$ extrapolation placing it near $1.2$. This limiting value is not universal (it is set by the lattice-to-continuum matching of the two harmonics over the Brillouin zone), but the locking itself is: An anisotropic start ($J_x \neq J_y$) flows to the same isotropic fixed point. The anisotropy is an irrelevant deformation absorbed by a rescaling of space, so that the two stress-tensor components must converge to their universal ratio regardless of the microscopic couplings.

\section{The fully irreducible vertex and a parquet benchmark}
\label{sec:parquet}

After determining the exact vertex through Monte Carlo simulations, we now turn our attention to its implications for diagrammatic theories. The vertex we have measured is irreducible in the particle-hole (\(ph\)) channel, but it does incorporate the resummation of the ladder diagrams in the two crossed channels: the particle-particle (\(pp\)) and the transverse particle-hole (\(\overline{\text{ph}}\)) channels. The fully irreducible vertex \(\Lambda\) is the key object in the parquet formalism, which is irreducible in all channels \cite{Diatlov1957,DeDominicis1964,Bickers1991}. Therefore, it is natural to consider what remains once these ladders are removed.

Since the field is real and the interaction involves a single on-site $\phi^4$ term, the connected four-point function maintains full crossing symmetry on the lattice (see App.~\ref{app:crossing}). The vertices from the three channels reduce to just one function, which is evaluated at three different momentum transfers: $q$ (ph), $k + p$ (pp), and $k - p$ ($\overline{\text{ph}}$). We have directly verified these identities on the connected correlator $\chi_c = \chi - \chi_0$ and found that they hold within the Monte Carlo data's error bars.

We construct the vertex function \(\Lambda\) from the same data using the parquet equation (our conventions for the parquet formalism are explained in App.~\ref{app:parquet}):
\begin{equation}
  \Lambda = \Gamma_{\text{ph}} + \Gamma_{\text{pp}} + \Gamma_{\overline{\text{ph}}} - 2F,
  \label{eq:lambda}
\end{equation}
where each vertex is obtained through a Bethe-Salpeter inversion, given by \(\Gamma_r = \chi_{0,r}^{-1} - \chi_r^{-1}\) for \(r \in (\text{pp}, \text{ph}, \overline{\text{ph}})\). Here, \(F = \chi_0^{-1}(\chi - \chi_0)\chi_0^{-1}\) represents the fully amputated vertex. 
In the case of a finite Euclidean lattice system, the Luttinger-Ward functional is proven to be unique~\cite{LinLindsey2018}. Consequently, the susceptibility \(\chi\) is positive definite and possesses the properties of a covariance. This implies that both \(\chi\) and each \(\Gamma_r\) are invertible, ensuring that \(\Lambda\) is finite by construction. 
It is worth noting that for fermionic systems, the Luttinger-Ward functional can become multivalued~\cite{Kozik2015, Schaefer2016, Gunnarsson2016, Gunnarsson2017}, which adds complexity to a similar analysis.

The locality of $\Lambda$ is illustrated in the inset of Fig.~\ref{fig:loc}. The relative-coordinate weight at one lattice spacing $(r_1)$ is only 0.2\% to 3\% of its on-site value $(r_0)$ across the entire window. We cannot discern the weight at larger distances due to noise. By stripping the Bethe-Salpeter ladders, we remove this tail as expected: $\Lambda$ is 20 to 110 times more local than the particle-hole vertex at $L=8$ and, within the resolution of our construction, behaves like a pure contact term.
This observation is not a result of finite-size artifacts. For the $L=16$ data at $\beta = 0.6$, we find that $\Lambda(r_1)/ \Lambda(r_0) = 0.0029 \pm 0.0030$, indicating a pure contact term within $1\sigma$. At $\beta = 0.64$, this ratio is $\Lambda(r_1)/ \Lambda(r_0) = 0.0150 \pm 0.0056$, which remains a pure contact term within $3\sigma$. At $\beta = 0.68$, we observe $\Lambda(r_1)/ \Lambda(r_0) = 0.0250 \pm 0.0064$, again a pure contact term within $4\sigma$. 
While it is likely that the fully irreducible vertex develops a short-ranged but low-amplitude contribution for $r \neq 0$ at very large system sizes, establishing this with certainty is challenging, even with high-precision Monte Carlo data. All local mean values remain consistent with the error bars observed for $L=8$.
Additionally, we note that removing the ladders also resolves the center-of-mass issue discussed in App.~\ref{app:caveat}. The center of mass (COM) structure of the 2PI vertex resides in the ladders, while $\Lambda$ carries less than 1\% of it, increasing only slightly as we approach $\beta_c$ in the $A_1$ channel. 
Our findings place the fRG/parquet expectations and methods, such as D$\Gamma$A (see App~\ref{app:dmft})~\cite{Toschi2007,Rohringer2018}, on solid ground for $\phi^4$ models.

\begin{figure}
  \includegraphics[width=\columnwidth]{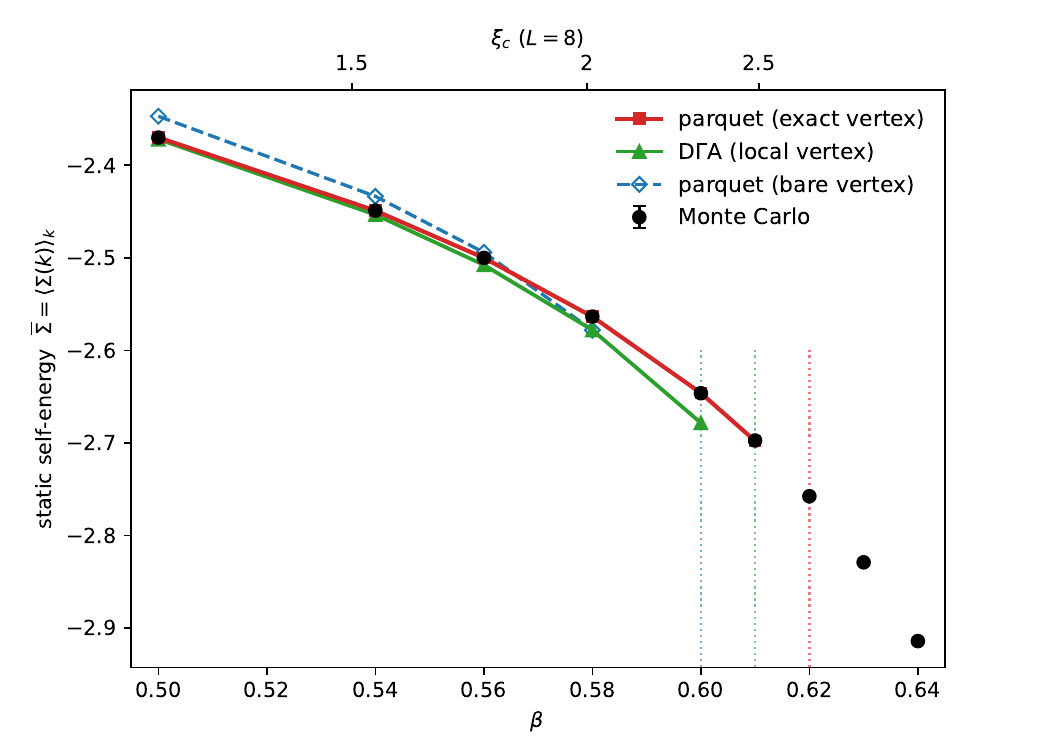}
  \caption{\label{fig:parquet}
  The local self-energy, denoted as \(\overline{\Sigma} = \langle \Sigma(k) \rangle_k\), is plotted against \(\beta\) for a system with \(L=8\). This data comes from a fully self-consistent solution using the parquet–Schwinger-Dyson–Dyson approach. The results show the full measured (exact) vertex represented by red squares, while the D$\Gamma$A local single-site vertex is indicated by green triangles. The blue open diamonds show the bare-vertex (parquet-approximation) result, \(\Lambda = -6\lambda/N\), included as a reference. These findings are compared to Monte Carlo simulations, with points that include bootstrap errors.
The exact vertex closely aligns with the Monte Carlo self-energy, whereas the local D$\Gamma$A vertex falls approximately 1\% short. Additionally, the D$\Gamma$A vertex is more strongly renormalized, which leads to a loss of stability at an earlier stage.  The dotted vertical lines indicate the points where each self-consistent iteration fails to converge, identified by Jacobian eigenvalues that exit the unit disk. It occurs first for the bare vertex, at \(\beta \simeq 0.60\), then for the D$\Gamma$A vertex at \(\beta \simeq 0.61\), and last for the exact vertex at \(\beta \simeq 0.62\). Both values are significantly below the critical point \(\beta_c \simeq 0.685\) and the finite-size critical value for \(L=8\), which is approximately \(\beta_c \approx 0.64(1)\). The upper x-axis displays the connected correlation length, as detailed in Appendix~\ref{app:fss}.}
\end{figure}

As the tail of \(\Lambda\) is obscured by noise, we must explore alternative methods to determine its relevance. To achieve this, we replace \(\Lambda\) with its effective contact value, defined as \(\lambda_{\text{eff}} = \frac{1}{N^2} \sum_{Q,k,p} \Lambda(k,p;Q)\), and then we close the parquet loop. It is worth noting that the term \(Q=0\) predominantly contributes to this summation; however, it exceeds \(\lambda_{\text{eff}}\) by an unacceptable 2-3\% when \(\beta \ge 0.6\), resulting in significantly poorer outcomes that are not presented here. We will proceed with our analysis in several steps.

\begin{figure}
  \includegraphics[width=\columnwidth]{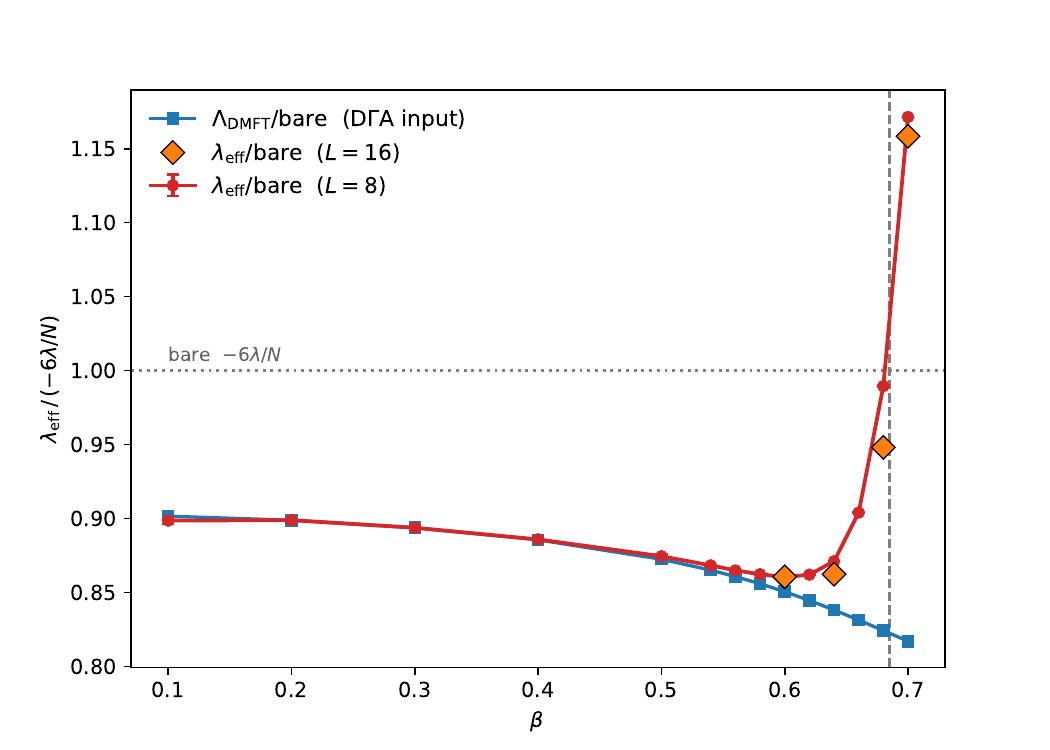}
  \caption{\label{fig:dmft}
  The ratio of the local fully irreducible vertex to the bare vertex as a function of \(\beta\) is shown, with the exact measured contact \(\lambda_{\rm eff}\) represented by red circles for \(L=8\) and orange diamonds for \(L=16\). The bootstrap errors are smaller than the symbols. These measurements are compared to the vertex \(\Lambda_{\rm DMFT}\) from a single-site dynamical mean-field theory, indicated by blue squares, as well as to the bare value, represented by a grey dotted horizontal line. The approximate location of the thermodynamic phase transition is indicated by a vertical grey dashed line at \(\beta_c \approx 0.685\). It is important to note that \(\lambda_{\rm eff}\) does not coincide with the bare vertex at \(\beta = 0\) because it reflects an atomic limit theory rather than a Gaussian free theory.
    }
\end{figure}

Firstly, as a preliminary test, we close the loop at the measured propagator. By inserting the single value $\lambda_{\rm eff}$ into the parquet approach at the Monte Carlo $G$ and propagating through Eq.~\eqref{eq:sde} and the Dyson equation, we can reproduce the measured self-energy to better than a tenth of a percent throughout the convergent window. This confirms that the physical solution obtained from Monte Carlo methods is a fixed point of the parquet formalism, even in the critical region (though excluding the thermodynamic phase transition).
From a physical perspective, the fully irreducible vertex functions as a screened local contact, which is only mildly renormalized from the bare interaction by a factor of approximately 0.85  (see Fig.~\ref{fig:dmft}) in the disordered phase, and this changes to an enhancement of approximately 1.15 at criticality. This indicates that the strong correlations contributing to the self-energy are primarily contained in the ladder diagrams that the parquet formalism resums, rather than in $\Lambda$ itself.

Secondly, our observation raises the natural question within the D$\Gamma$A framework: how does the measured contact compare with the local vertex that a practitioner would actually use, specifically the fully irreducible vertex of a dynamical mean-field impurity? In classical $\phi^4$ field theory, the impurity corresponds to a single site, which is solved self-consistently within a Gaussian bath through quadrature (this approach is equivalent to the DMFT approximation). The fully irreducible vertex, denoted as $\Lambda_{\rm DMFT}$, is derived from Eq.~\eqref{eq:lambda} specialized for one mode (refer to App.~\ref{app:dmft} for a detailed derivation). This leads to the D$\Gamma$A approximation, and its vertex is compared to $\lambda_{\rm eff}$ in Fig.~\ref{fig:dmft}. It is important to note that due to the simplicity of the theory, the fully irreducible vertex is frequency-independent, which is why we denote it as $\Lambda_{\rm DMFT}$.

Figure~\ref{fig:parquet} compares the local self-energy to the one obtained through $\lambda_{\rm eff}$. This provides a fair test because local theories, such as DMFT and D$\Gamma$A, are expected to effectively capture local quantities. However, there is a notable difference compared to the previous paragraph, which analyzed the parquet equation for the measured propagator: here, we need to solve the full parquet equations (known as closure) without prior knowledge of the exact $G$. This can lead to stability issues (see below); in fact, the physical solution transitions from an attractive fixed point to a repelling one, notably throughout the entire critical region. For future reference, we refer to this procedure as a parquet–Schwinger-Dyson–Dyson sweep, or PSD; the earlier approach of parquet–Schwinger-Dyson sweeps at a fixed, exact $G$ will be referred to as PS.

\begin{figure}
  \includegraphics[width=\columnwidth]{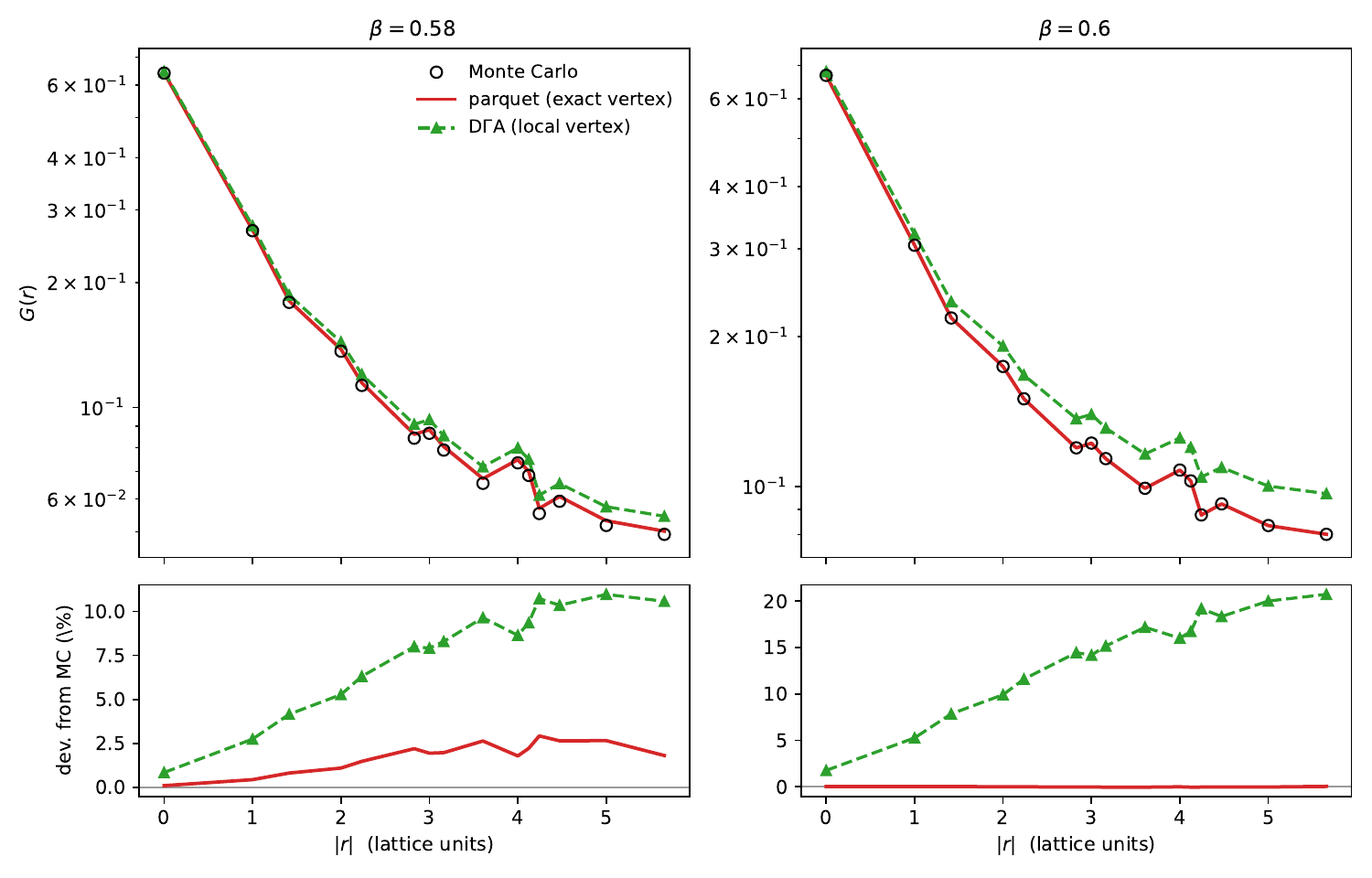}
  \caption{\label{fig:gr}
  The real-space propagator \( G(r) \) is compared between the fully self-consistent parquet solution—derived from both the fully measured (statistically exact) vertex and the approximate D$\Gamma$A local vertex—and the Monte Carlo propagator (also statistically exact) at \(\beta = 0.58\) and \(0.60\) (with \(L = 8\), on a logarithmic scale). The exact vertex accurately reproduces \( G(r) \) at all distances, while the local D$\Gamma$A vertex shows an overshoot at large \(|r|\). In the lower panels, the percentage deviation from the Monte Carlo results is presented. Statistical errors, obtained through bootstrapping, are approximately \(0.1\%\), which is smaller than the plotted symbols. }
\end{figure}

Nevertheless, the figure clearly shows that away from criticality, $\Lambda_{\rm DMFT}$ and $\lambda_{\rm eff}$ agree within a few percent. Thus, we validate the assumption that the fully irreducible vertex is local and can be well approximated by a local impurity against the exact vertex. However, as we approach $\beta_c$, the measured contact is enhanced while the D$\Gamma$A vertex remains nearly constant (and incorrectly always screens). 
Next, we examine the non-local properties, which are not directly controlled by DMFT/D$\Gamma$A. As long as the (connected) correlation length $\xi_c$ remains less than one lattice spacing, DMFT/D$\Gamma$A remains virtually exact. However, as $\xi_c$ increases, the limitations of the local vertex become evident, as illustrated in Fig.~\ref{fig:gr}. While the correlation function $G(r)$ obtained with the exact vertex aligns with Monte Carlo results at all distances, the D$\Gamma$A local vertex overshoots the long-range propagator by up to 10–20\%, and this deviation increases with the distance $|r|$. This discrepancy highlights the non-local corrections that a momentum-independent $\Lambda$ cannot accommodate.

Thirdly, we return to the issue of convergence. The loss of convergence observed in the parquet self-consistency equations is not attributed to a physical singularity; rather, it is a result of damped fixed-point iterations acquiring a Jacobian eigenvalue of unit modulus or greater. This causes the physical fixed point to become repulsive~\cite{Essl2026}.

This picture is supported by the Monte Carlo data: when inserting the exact solution at a fixed propagator, the spectral radius of the map increases smoothly from \(0.90\) at \(\beta=0.60\) to \(0.94\) at \(\beta=0.64\), crossing unity between \(\beta=0.64\) and \(0.66\). This places the linear stability threshold at approximately \(\beta\simeq0.65\). The self-consistent iteration, which couples this map to the Dyson update, loses convergence significantly earlier, at \(\beta\simeq0.62\), because the Dyson feedback is itself destabilizing (see below and App.~\ref{app:signflip}).
Anderson-accelerated iterations~\cite{WalkerHomerNi,FangSaad2009} and a \(\beta\)-annealed continuation do not help; they seek an attractor and stop at the same point where stability is lost. However, a Jacobian-free Newton-Krylov root-finder~\cite{Knoll2004}, which can converge to repulsive fixed points, does manage to extend the solution up to \(\beta=0.65\)—the onset of the single-mode repeller—but stalls just above this point at the mode proliferation phase (\(\beta\ge0.66\);  see App.~\ref{app:signflip}). Therefore, the extent to which such methods can be applied to the two-dimensional lattice up to \(\beta_c\) remains an open question.
The scenario presented in Ref.~\cite{Essl2026} suggests a procedure to stabilize the equations in the non-convergent regime, particularly when dealing with a large eigenvalue. We reference App.~\ref{app:signflip}, where we analyze this in detail, but summarize the two key results here: First, at its onset, there is a \emph{single} mode (specifically, the \(A_1\) zero-transfer (order-parameter) direction), so the fixed-\(G\) (PS) map remains an attractor up to its wall (\(\beta\approx0.64\)–\(0.65\)). However, closing the Dyson loop (PSD) feeds that same soft mode back through the propagator, which makes it repulsive earlier (Eq.~\eqref{eq:mprime}): self-consistency is \emph{destabilizing}, and the PSD map fails first. 
Second, this wall is not a fixed barrier below \(\beta_c\); rather, its onset is a finite-size feature that recedes toward \(\beta_c\) with increasing system size. Near criticality, the unstable set proliferates into many complex modes, and single-mode stabilization is no longer sufficient (App.~\ref{app:signflip}).
In summary, solving the parquet equations remains a notoriously challenging and open problem, even with knowledge of the physical irreducible vertex. We emphasize again that this is not a limitation of the parquet equations themselves; they remain valid in the critical region.

\section{Discussion}
We have reconstructed the 2PI vertex for lattice \(\phi^4\) theory using Monte Carlo measurements of the connected two-particle correlator, which can serve as a benchmark object against which various theories such as fRG, parquet, D\(\Gamma\)A, and Schwinger–Dyson can be tested. We then analyzed the 2D \(\phi^4\) Ising transition from the vertex perspective. Our main results can be summarized as follows:

First (see Figs.~\ref{fig:irrep} and~\ref{fig:qdep}), the soft sector at the transition is multidimensional. The \(A_1\) irreducible representation drives the instability, but the \(B_1\) and \(B_2\) representations cooperate and strengthen as the system size increases. Consequently, the minimal reduced description of the near-critical vertex spans these channels rather than relying solely on the \(A_1\) mode. The contribution from \(B_2\) is tied to \(B_1\) because of the emergent rotational invariance at the critical point.

Second, the two vertices differ in range (see Figs.~\ref{fig:loc} and~\ref{fig:dmft}). The fully irreducible vertex \(\Lambda\) is spatially compact—a near-contact object with approximately 40 times more amplitude at \(r=0\) than at \(r=1\) even at criticality—while the particle-hole irreducible (2PI) vertex develops a power-law tail at criticality and shows non-monotonic behavior at small \(r\) in the relative coordinate, along with a complex structure in the center-of-mass coordinate. In the disordered phase, both vertices are localized within \(r \approx 1\); here, \(\Lambda\) screens the bare contact by approximately 10\%, whereas toward criticality, the effective contact is instead enhanced (as shown in Fig.~\ref{fig:dmft}).

Third, the D\(\Gamma\)A approximation for \(\phi^4\) theory accurately reproduces the fully irreducible vertex and the local self-energy (see Fig.~\ref{fig:parquet}), but begins to fail when the correlation length exceeds one lattice spacing. This failure is first evident in the closure of the correlation function (see Fig.~\ref{fig:gr}) and can also be observed, albeit to a lesser extent, in local quantities (see Fig.~\ref{fig:parquet}).

Fourth, in the critical region, the physical solution of the parquet-SDE-Dyson (PSD) equations turns into a repulsive fixed point, and we clarify the structure of this instability (see App~\ref{app:signflip}). This instability begins as a single order-parameter (\(A_1\)) mode, which contributes an eigenvalue to the Jacobian with a magnitude greater than one. The fixed-propagator map (PS) remains an attractor up to a certain boundary that self-consistency (PSD) pushes to a lower wall. This boundary is a finite-size feature that approaches \(\beta_c\) as the system size increases. Following Ref.~\cite{Essl2026}, we stabilize that single mode, enhancing convergence; however, nearer to the transition, the unstable set proliferates into many largely complex modes, and a rank-one flip is no longer sufficient. This is a stability, not an existence, issue—it is not a property of the PSD equations themselves, nor is it related to the uniqueness of the Luttinger-Ward functional (which is guaranteed to be unique in our case); only the solver's Jacobian becomes unstable.

Collectively, these results establish the directly measured vertex as a controlled, statistically valid benchmark for two-particle-based theories—such as parquet, D\(\Gamma\)A, fRG, and Schwinger–Dyson—on the disordered side of lattice field theory, and they precisely locate where these closures begin to fail.

In principle, this approach could be extended to any \(N\)-component \(\phi^4\) theory on any \(D\)-dimensional finite lattice in the disordered phase, provided an efficient cluster algorithm exists for the two- and four-point correlators. However, we caution that this is not a scalable, general-purpose method: measuring the fully connected four-point function accurately enough to extract the Bethe–Salpeter ladders is costly in both statistics and memory, and these costs grow rapidly with the number of components, dimensionality, and correlation length. Therefore, we expect this method to remain practical only for a few comparatively simple models. The most natural candidates include the 3D Ising model, the \(O(N)\) vector models (with the \(N=2\) quasi-long-range-ordered critical phase being a particularly challenging test), and Heisenberg \(O(3)\) spin models, whose vertex already couples nearest neighbors. In each case, the disordered side is accessible, but solving the closure relation (PSD) near criticality remains a formidable and likely model-dependent problem; we see no reason to expect it to be easier in that region than it was in our case. Future work will address the ordered phase and extensions to bosonic and fermionic many-body systems, where the vertex is significantly more complex and further conceptual complications arise, including the multivalued nature of the Luttinger-Ward functional in the fermionic case.

\section{Acknowledgements}
\label{sec:ack} 
\noindent \emph{Thanks --}
I wish to thank Lei Wang for an initial conversation on recent AI developments and drawing my attention to Ref.~\cite{LinLindsey2018}. \\
\emph{Funding --}
We acknowledge financial support from the Deutsche Forschungsgemeinschaft (DFG), project Nr 530111096.
The project/research is part of the Munich Quantum Valley, which is supported by the Bavarian state government with funds from the Hightech Agenda Bayern Plus.\\
\emph{ Data Analysis and Code Generation Disclosure -- }
This project was the author's first attempt to make extensive use of the current generation of AI large-language models (LLM). The mathematical derivations, numerical simulations, and linear algebra stabilization routines detailed were developed with the computational assistance of, primarily, Anthropic Claude 4.8 Opus and, to a lesser extent, OpenAI ChatGPT-5.5. Specifically, the Anthropic AI tool was used to draft Python scripts for solving the parquet and $D \Gamma A$ self-consistency equations and adding observables to the (primarily human-written) Monte Carlo codes, and to optimize the statistical analysis pipeline. All mathematical outputs, code logic, and data matrices were subjected to rigorous human supervision, manual derivation cross-checks, and independent verification by the author to guarantee accuracy and physical validity to the best of their knowledge and skill. The OpenAI tool was used as a conversation tool about the progress of the project. Google Gemini was used for broad literature searches and minor scripting. The author claims full responsibility. The paper text has been extensively edited by a human and then checked by Grammarly. \\
\emph{Data availability -- }
The ALPSCore libraries were used in the Monte Carlo simulations~\cite{ALPSCore1,ALPSCore2}.
Data and certain programs used in this work will be made publicly available under \url{https://github.com/LodePollet} upon publication and earlier upon reasonable non-anonymous request.

\appendix

\section{Conventions and definitions}
\label{app:conv}

To enhance the reproducibility of our results, we provide a collection of definitions, conventions, and normalizations. 
All momenta run over the discrete Brillouin zone of the $L \times L$ lattice, represented as $k = \tfrac{2\pi}{L}(n_x, n_y)$, where $N = L^2$.

\emph{Fields and propagator.} The Fourier transform is symmetric,
$\phi_k=N^{-1/2}\sum_x e^{-ik\cdot x}\phi_x$, so that the fact that  the fields are real implies 
$\phi_{-k}=\phi_k^{*}$. The propagator is $G(k)=\langle|\phi_k|^2\rangle$. With the action
$S=-\beta\sum_{\langle ij\rangle}\phi_i\phi_j+m^2\sum_i\phi_i^2+\lambda\sum_i\phi_i^4$ the free inverse propagator is
\begin{equation}
  G_0^{-1}(k)=2\bigl(m^2-\beta(\cos k_x+\cos k_y)\bigr),
  \label{eq:g0}
\end{equation}
and we work at $m^2=0$, $\lambda=\tfrac12$. In the $(1/n!)V_n$ convention the bare amputated quartic
vertex is $V_4=24\lambda$; the symmetric transform turns the on-site $\lambda\sum_x\phi_x^4$ into a
momentum-conserving vertex $V_4^{\rm mom}=24\lambda/N$.

\emph{Two-particle correlator and particle--hole vertex.} With
$\rho_q(k)=\phi_k^{*}\phi_{k+q}$ the measured connected correlator and its non-interacting counterpart
are
\begin{align}
  \chi(k,p;q) &= \langle\rho_q(k)\rho_q(p)^{*}\rangle-\langle\rho_q(k)\rangle\langle\rho_q(p)^{*}\rangle,\\
  \chi_0(k,p;q) &= G(k)G(k+q)\,[\delta_{k,p}+\delta_{p,-k-q}],
  \label{eq:chi0}
\end{align}
where the exchange term $\delta_{p,-k-q}$ is required by $\phi_{-k}=\phi_k^{*}$. The particle--hole 2PI
vertex, the connected part, and the full amputated vertex are
\begin{align}
  \Gamma_{\rm ph}&=\chi_0^{-1}-\chi^{-1}, &
  \chi_c&=\chi-\chi_0, &
  F&=\chi_0^{-1}\chi_c\,\chi_0^{-1},
  \label{eq:GammaF}
\end{align}
and the instability is monitored through $K=\chi_0^{1/2}\Gamma_{\rm ph}\chi_0^{1/2}$, whose leading
eigenvalue reaches unity at the Thouless point (physical phase transition). Throughout, matrix operations act on the $(k,p)$
indices at fixed transfer $q$ on the physical (non-null) subspace of $\chi$.

\section{Remark on the locality of the 2PI vertex}
\label{app:caveat}
A remark on the interpretation of Fig.~\ref{fig:loc} is in order here. The relative-coordinate kernel $\Gamma_{\mathrm{rel}}(r)$ is obtained by summing over all $(k,p)$ at fixed $k-p$, {\it ie}, by averaging the vertex over the center-of-mass momentum $k+p$. It hence measures the range of the convolution part of the vertex and is blind to any dependence on $k+p$. 
This distinction matters because the center-of-mass part is directly measurable as $\Gamma_{\mathrm{com}}=\Gamma-\Gamma_{\mathrm{conv}}$ with $\Gamma_{\mathrm{conv}}(k,p)=g(k-p)$, $g(d)=\langle\Gamma(k,k-d)\rangle_k$, and we quantify its weight $\lVert\Gamma_{\mathrm{com}}\rVert^2/\lVert\Gamma\rVert^2$ on the physical inversion-even subspace.
Deep in the disordered phase this weight is negligible: it is below $0.1\%$ for $\beta\le0.4$.
On approach to the transition however it grows rapidly and monotonically, reaching $\sim 4\%$ at $\beta=0.60$, $\sim 15\%$ at $0.64$, and $\sim 40\%$ of the total vertex norm by $\beta=0.68$: Near criticality, the center-of-mass structure is comparable to the convolution part. It is moreover organized in precisely the soft sector that drives the instability (Fig.~\ref{fig:irrep}): decomposing  into $C_{4v}$ irreducible representations, the weight  at $\beta=0.68$ is carried foremost by $A_1$ (energy, $26.7\%$), then by $B_1$ and $B_2$ (the stress-tensor pair, $8.8\%$ and $2.5\%$, respectively), with $A_2$ negligible ($0.7\%$), and this holds for both the $L=8$ and $L=16$ data sets. The near-locality of $\Gamma_{\mathrm{rel}}$ is therefore a partial statement: the relative-coordinate kernel stays short ranged (Fig.~\ref{fig:loc}), while a second, center-of-mass structure emerges and grows toward half the vertex as $\beta_c$ is approached. 
Note that this is not the case for the fully irreducible vertex discussed below: the COM structure lives almost exclusively in the ladders.

\section{Crossing and the fully irreducible vertex}
\label{app:crossing}

Because the field is real and the interaction is a single local $\phi^4$ term, the connected four-point
function is fully crossing symmetric. Therefore, the two crossing relations
\begin{equation}
  \chi_c(k,p;q)=\chi_c(k,k+q;p-k)=\chi_c(k,p;-k-p-q)
  \label{eq:crossing}
\end{equation}
hold identically (and we have verified this on the data); they map the particle-hole
transfer $q$ onto the transverse ($\overline{\rm ph}$, transfer $k-p$) and particle-particle (pp,
transfer $k+p$) channels. Each channel carries its own vertex
$\Gamma_r=\chi_{0,r}^{-1}-\chi_r^{-1}$, $r \in \{{\rm ph}, {\rm pp}, \overline{\rm ph}\}$, obtained from the same measured function evaluated at the crossed
transfer. The fully irreducible vertex follows from the parquet equation
[cf.\ Eq.~\eqref{eq:lambda}]
\begin{equation}
  \Lambda=\Gamma_{\rm ph}+\Gamma_{\rm pp}+\Gamma_{\overline{\rm ph}}-2F .
  \label{eq:lambda-app}
\end{equation}
Since each channel-irreducible vertex satisfies $\Gamma_r = F - \Phi_r$, summing over the three channels gives $\sum_r \Gamma_r = 3F - \sum_r \Phi_r$. Using the parquet identity $F = \Lambda + \sum_r \Phi_r$, i.e.\ $\sum_r \Phi_r = F - \Lambda$, immediately gives Eq.~\eqref{eq:lambda-app}.
Through the exchange structure of Eq.~\eqref{eq:chi0} we see that $\chi_0$ acts as
$2\,G(k)G(k+q)$ on the symmetric physical subspace (the factor of 2 is a consequence of the bosonic (real-field) counting of the reducible
diagrams~\cite{Eckhardt2023}), so that each $\chi_0^{-1}$ in $F$ carries a factor
$\tfrac12$.
There is also a factor $\tfrac14$ between the amputated bare vertex $V_4=24\lambda$ and the contact measured in our normalization,
and this places the amputated tree vertex at $F^{\rm tree}=-V_4/4=-6\lambda/N$.  This fixes
the normalization convention used throughout the paper in which the local contact $\lambda_{\rm eff}=N^{-2}\sum_{Q,k,p}\Lambda(k,p;Q)$ and
the bare parquet input ($-6\lambda/N$) are quoted.

\section{Parquet self-consistency and the self-energy}
\label{app:parquet}
Whereas App.~\ref{app:crossing} extracts $\Lambda$ from the measured vertices, here we solve the parquet equations with a given $\Lambda$.
We have followed the functional derivation of the parquet equations for $\phi^4$~\cite{Eckhardt2023}.
As stated in App.~\ref{app:crossing}, because $F$ is crossing symmetric, the three reducible parts (BSE equations) are images of a single function
$\Phi_{\rm ph}[Q]$, and the parquet equations collapse to a fixed-point iteration over the transfer
$Q$ alone,
\begin{align}
  \Gamma_{\rm ph}[Q](a,b)&=\Lambda+\Phi_{\rm ph}(a,a+Q;b-a)  \nonumber \\
  {} &  +\Phi_{\rm ph}(a,b;-(a{+}b{+}Q)),\\
  F[Q]&=\bigl(1-\Gamma_{\rm ph}[Q]\,\chi_0[Q]\bigr)^{-1}\Gamma_{\rm ph}[Q],\\
  \Phi_{\rm ph}[Q]&=F[Q]-\Gamma_{\rm ph}[Q].
  \label{eq:parquet}
\end{align}
The self-energy follows from the Schwinger-Dyson equation as $\Sigma(k) = \Sigma_H(k) + \Sigma_{\sunsetdiag}$ with $\Sigma_H =  -12\lambda\,\langle\phi^2\rangle$ the  Hartree term and the sunset diagram~\cite{BickersScalapino1992} given by
\begin{equation}
  \Sigma_{\sunsetdiag}(k) = -\;\frac{16\lambda}{N}\!\sum_{k_2,k_3}\! G(k_2)G(k_3)G(k_4)\,F(k,k_2,k_3,k_4),
  \label{eq:sde}
\end{equation}
with $k_4=-k-k_2-k_3$, $\langle\phi^2\rangle=N^{-1}\sum_k G(k)$, and the full vertex entered in
particle-hole form, $F(k,k_2,k_3,k_4)=F[Q{=}k_2{+}k](-k,k_3)$. 
The coefficient of the sunset is fixed: the field-theory sunset carries $-4\lambda/N$ acting on the field-normalized vertex, and the measured
$F=F^{\rm mom}/4$ of App.~\ref{app:crossing} converts this to $-16\lambda/N$.  Equivalently,
Eq.~\eqref{eq:sde} with a constant $F=-6\lambda/N$ reproduces the direct second-order sunset
$96\,\lambda^2 N^{-2}\sum GGG$ identically. Inserting the \emph{measured} $F$ into Eq.~\eqref{eq:sde}
reproduces the Monte Carlo self-energy over the full Brillouin zone to $4\times10^{-4}$ and within error bars. The propagator closes
through the Dyson equation $G(k)=[G_0^{-1}(k)-\Sigma(k)]^{-1}$.

This set of equations is iterated, in principle, until $\Phi_{\rm ph}$ converges although the stability of the parquet equations is difficult to guarantee (see main text and App.~\ref{app:signflip}).

\section{Single-site dynamical mean-field theory (DMFT) and $D\Gamma A$}
\label{app:dmft}

The impurity of Fig.~\ref{fig:dmft} is one site in a self-consistent Gaussian bath, with static Weiss
field $a$ and action $S_{\rm imp}=\tfrac12 a\,\phi^2+\lambda\phi^4$; its moments
$\langle\phi^2\rangle_{\rm imp}$ and $\langle\phi^4\rangle_{\rm imp}$ are single integrals. The
self-consistency loop, in the normalization of Eq.~\eqref{eq:g0}, is
\begin{equation}
  G_{\rm loc}=\Bigl\langle\tfrac{1}{G_0^{-1}(k)-\Sigma}\Bigr\rangle_k,\quad
  a=G_{\rm loc}^{-1}+\Sigma,\quad
  \Sigma=a-\tfrac{1}{\langle\phi^2\rangle_{\rm imp}},
  \label{eq:dmft}
\end{equation}
iterated to $\langle\phi^2\rangle_{\rm imp}=G_{\rm loc}$. 
This constitutes the $\phi^4$ equivalent of dynamical mean-field theory (DMFT). Its extension to $D \Gamma A$, which includes the local vertex, is as follows:
For a single mode the three channels of
Eq.~\eqref{eq:lambda-app} collapse to one scalar: with $\chi_0=2\langle\phi^2\rangle_{\rm imp}^2$ and
$\chi=\langle\phi^4\rangle_{\rm imp}-\langle\phi^2\rangle_{\rm imp}^2$,
\begin{equation}
  \Gamma=\chi_0^{-1}-\chi^{-1},\quad F=\chi\chi_0^{-2}-\chi_0^{-1},\quad
  \Lambda_{\rm D \Gamma A}=\frac{3\Gamma-2F}{N},
  \label{eq:lamdmft}
\end{equation}
the final division by $N$ bringing the single-site vertex in line with the lattice (momentum) normalization of
$\lambda_{\rm eff}$.

\section{Conditioning, noise, and inversion}
\label{app:stab}

Through Monte Carlo simulations we have access to a stochastically sampled 4-point (and 2-point) correlator. 
The inversion of such a stochastically sampled correlator is
rank deficient. The stability of the procedure rests on separating the exactly null,
the well-measured, and the noise-limited parts of $\chi$, and on choosing the metric in which each
quantity is read.

\emph{Null space and physical subspace.} The relation expressing that the field is real, $\rho_q(k)=\rho_{-q}(-k)$,  makes $\chi$ exactly rank deficient: roughly half of its eigenvalues vanish
identically ($34$ of $64$ at $L=8$; $130$ of $256$ at $L=16$). These exact zeros  must be
projected out before any inversion.
After this projection the condition number of the full matrix is reduced to $\sim\!10^{3}$--$10^{4}$. 
All inversions in Eqs.~\eqref{eq:GammaF} are pseudo-inverses restricted to
this subspace, defined by a relative truncation threshold $\tau$: an eigen-direction of the matrix is
kept only if its eigenvalue exceeds $\tau$ times the largest eigenvalue. We use $\tau\simeq10^{-6}$ and find the
results stable across the range $\tau\in[10^{-6},10^{-4}]$. The bare bubble $\chi_0$ of Eq.~\eqref{eq:chi0} is, by contrast, block diagonal in
the pairs $(k,-k-q)$ and is inverted exactly, so $\chi_0^{-1}$ and $\chi_0^{1/2}$ introduce no error.

\emph{Robustness through the metric.} The leading eigenvalue of the symmetrised kernel
$K=\chi_0^{1/2}\Gamma\chi_0^{1/2}$ is determined by the soft high-susceptibility collective modes (the $A_1$
energy mode and the $B_1,B_2$ stress modes). Those are well sampled, enabling a precise determination of the leading eigenvalues.
The real-space vertex $\Gamma_{\rm rel}(r)$ needs the small eigenvalues of $\chi$ and is therefore far noisier.
One must rank the spectrum of $K$ by its \emph{positive} eigenvalues: large
negative eigenvalues are pseudo-inverse noise from the near-null sector, not physical, and ranking by
$|\cdot|$ returns nonsense. The same $\chi_0^{1/2}$ weighting suppresses a purely numerical artifact
that appears near criticality. As the soft mode grows, the truncation threshold
$\tau\,\lambda_{\max}(\chi)$ (and this is here the susceptibility $\lambda_{\max}(\chi)=\chi(0)=G(0)$ ) can regularize by dropping a small $\chi$ eigenvalue at  $M=(\pi,\pi)$:  although the large $\chi_0^{-1}(M,M)$ (recall $\chi_0(M)=G(M)G(M+q)$ is tiny there)
survives and produces a spurious spike in the raw $\Gamma = \chi_0^{-1} - \chi^{-1}$, this spurious peak is absent from $K$ as can be seen from writing $K$ as 
$K=\mathbf{1}-\chi_0^{1/2}\chi^{-1}\chi_0^{1/2}$ in which the residual
$\chi_0^{1/2}\chi^{-1}\chi_0^{1/2}$ carries only the small $\chi_0(M)$ and is suppressed rather than
amplified.
Where the raw vertex itself is needed, it is removed by an absolute-scale rather than relative
regularisation.

\emph{The fully irreducible vertex is a difference of large terms.} Equation~\eqref{eq:lambda-app}
subtracts quantities of comparable magnitude, so it amplifies the sampling noise on the small matrix
elements.
We therefore symmetrize $\Lambda$ over the $C_{4v}$ group elements, which is
physically exact. What survives cleanly is the contact: the $r=0$
amplitude $\lambda_{\rm eff}=N^{-2}\sum_{Q,k,p}\Lambda(k,p;Q)$ is orthogonal to the noisy tail and is
determined to a precision of $\sim\!0.3\%$ whereas extracting a range $\xi_\Lambda$ 
from the tail is not meaningful on the present statistics. 

\emph{Parquet iteration.} The Bethe-Salpeter step of Eq.~\eqref{eq:parquet} is solved as a linear
least-squares problem for $(1-\Gamma\chi_0)F=\Gamma$ rather than by explicit inversion, which is stable
even as $1-\Gamma\chi_0$ becomes singular. The iteration is started from the Monte
Carlo propagator, under-relaxed (mixing $0.25$), and guarded by a divergence test
($\max|\Phi_{\rm ph}|>10^{6}$). The divergence encountered for $\beta\gtrsim0.64$ is the
fixed-point-iteration instability discussed in Sec.~\ref{sec:parquet}. 

\section{Conventional finite-size scaling analysis}
\label{app:fss}
\begin{figure*}
  \includegraphics[width=\textwidth]{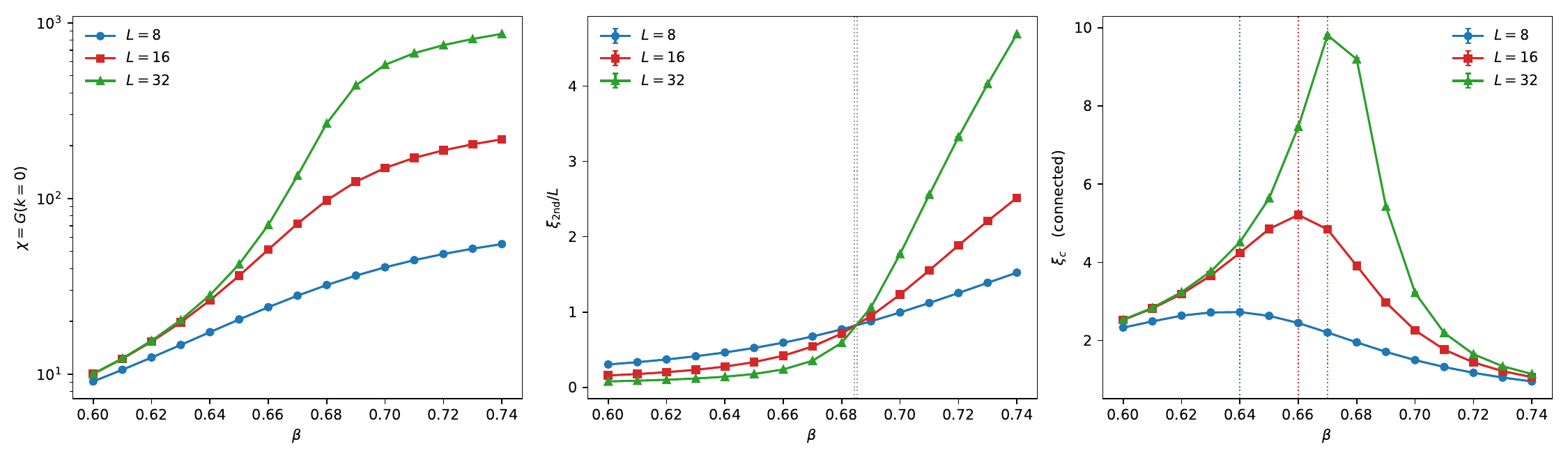}
  \caption{\label{fig:fss}
  Conventional finite-size scaling at $L=8,16,32$. Left: zero-momentum susceptibility $\chi=G(k{=}0)$,
  monotonic. Centre: second-moment ratio $\xi_{\rm 2nd}/L$, whose
  crossing (dotted) sits at $\beta_c\simeq0.685$ in a nearly $L$-independent way. Vertical lines are located at the crossings between ($L=8, L=16$) and  ($L=16, L=32$). Right: ratio of the connected correlation length to system size
  $\xi_c/L$; its peak and crossings drift upward with $L$, overshooting the second-moment crossing (this estimator peaks just beyond the transition; see text). Error bars are obtained through bootstrapping. Vertical lines correspond to peak positions on the $\beta$ grid.  }
\end{figure*}

As an independent check on the location of the transition we perform a standard finite-size-scaling
analysis of the propagator at $L=8,16,32$ (Fig.~\ref{fig:fss}). The zero-momentum susceptibility
$\chi=G(k{=}0)$ rises monotonically: on a finite lattice the raw $G(0)$
acquires the magnetization (order parameter) above $\beta_c$. The
Binder-cumulant analogue here is the renormalization-group-invariant ratio of the correlation length to the system size $\xi/L$. The second-moment correlation length $\xi_{\rm 2nd}$~\cite{Cooper1982},
\begin{equation}
\xi_{\rm 2nd} = \frac{1}{2\sin(\pi / L) } \sqrt { \frac{ G(0) }{ G(k_1)} -1}, \quad k_n = \left( \frac{2 n\pi}{L},0 \right),
\end{equation}
(where $G(k_1)$ is averaged over the two axis directions) gives a crossing that is almost $L$-independent, $\beta_c\simeq0.685$ for both the $(8,16)$ and
$(16,32)$ pairs. The second moment is however built on $G(0)$ and is mildly condensate-biased near the transition (and also much larger than $L$ on the ordered side for the same reason).
It is therefore only meaningful on the disordered side. 
The connected correlation length $\xi_c$ is measured from the effective mass at the two
lowest nonzero momenta and is free of the magnetization. It is computed as
\begin{equation}
\xi_{c}^{-2} =   \frac{ \hat{k}^2(k_2) G(k_2) - \hat{k}^2 G(k_1)}{G(k_1) - G(k_2)},
\end{equation}
where $\hat k^2(k)=2\bigl[(1-\cos k_x)+(1-\cos k_y)\bigr]$.
The connected correlation length is sharper than the second-moment correlation length: $\xi_c/L$ peaks at a pseudo-critical
coupling that drifts upward with system size, $\beta^\ast(L)\simeq0.64,\,0.66,\,0.67$ at $L=8,16,32$,
and its crossings move up in step, $0.656\to0.670$ from $(8,16)$ to $(16,32)$. Both the peak and the
crossing drift upward and, extrapolated linearly in $1/L$, land somewhat above the clean second-moment crossing; the same offset appears in the
vertex signal, whose $A_1$ eigenvalue (Fig.~\ref{fig:irrep}) peaks at $\beta\simeq0.70$ with little size
dependence. As discussed in Sec.~\ref{sec:vertexphys}, these estimators peak just \emph{beyond} the transition rather than at it, so we take the second-moment crossing, $\beta_c\simeq0.685$ (panel (b)), as the thermodynamic value. Note that this quantity never equals the full (or even half) the system size. The bootstrap uncertainty on each crossing is below $10^{-3}$; the spread quoted here is the
systematic difference between estimators and the finite-size drift, not statistical noise.

\section{Rank-one sign-flip stabilization at the edge of the convergence window}
\label{app:signflip}
Throughout this appendix we focus on the properties (Jacobian) of the numerical fixed-point map used to solve the parquet-SDE and parquet-SDE-Dyson equations.

Let us recap what we observe numerically for $L=8$. The finite-size transition point is at $\beta=0.64-0.65$ (see Fig.~\ref{fig:fss}(c)), the thermodynamic one at $\beta_c \approx 0.685$ (see Fig.~\ref{fig:fss}(b)).
The parquet-SDE-Dyson (PSD) equations could be solved by damped fixed point iterations up to $\beta=0.62$ (Fig.~\ref{fig:parquet}). Keeping $G$ fixed at its exact Monte Carlo value, the parquet-SDE (PS) equations could be solved by damped fixed-point iterations up to $\beta=0.64$. For $\beta > 0.62$ the physical solution corresponds to a repulsive fixed point in the PSD equations. The fixed-$G$ PS map remains an attractor up to $\beta=0.64$ (so damped iteration already converges there), and only turns repulsive at $\beta=0.65$; a Jacobian-free Newton-Krylov root-finder, which can reach repulsive fixed points, therefore extends the PS solution to $\beta=0.65$ before stalling at the mode proliferation ($\beta\ge0.66$).

That the PSD wall lies \emph{below} the PS wall is a direct consequence of the Dyson closure being destabilizing. With $G$ frozen (PS), the sensitivity of the map along the $A_1$ soft mode is set by the ladder resummation, $\sim 1/(1-\lambda_{A_1})$. Closing the Dyson loop (PSD) feeds the self-energy shift back into the propagator, and hence back into the vertex, adding a crossed contribution to the Jacobian eigenvalue along that same mode,
\begin{equation}
  M'_{A_1} \;=\; \Big[\frac{1}{(1-\lambda_{A_1})^2}-1\Big]\,c ,\qquad c\simeq 0.27 ,
  \label{eq:mprime}
\end{equation}
where $c$ is the (crossed) propagator-feedback weight measured at the wall. This term is positive and grows faster than the PS sensitivity as $\lambda_{A_1}\to1$, so it drives the self-consistent eigenvalue through unity at a \emph{smaller} $\lambda_{A_1}$---and hence a lower $\beta$---than the fixed-$G$ map. The physical vertex is thus an attractor of the PS map up to its (higher) wall, whereas adding self-consistency turns the very same order-parameter mode repulsive earlier: the PSD map fails first. The fixed point itself is unchanged; only the Jacobian of the closure differs.

\begin{figure}
  \includegraphics[width=\columnwidth]{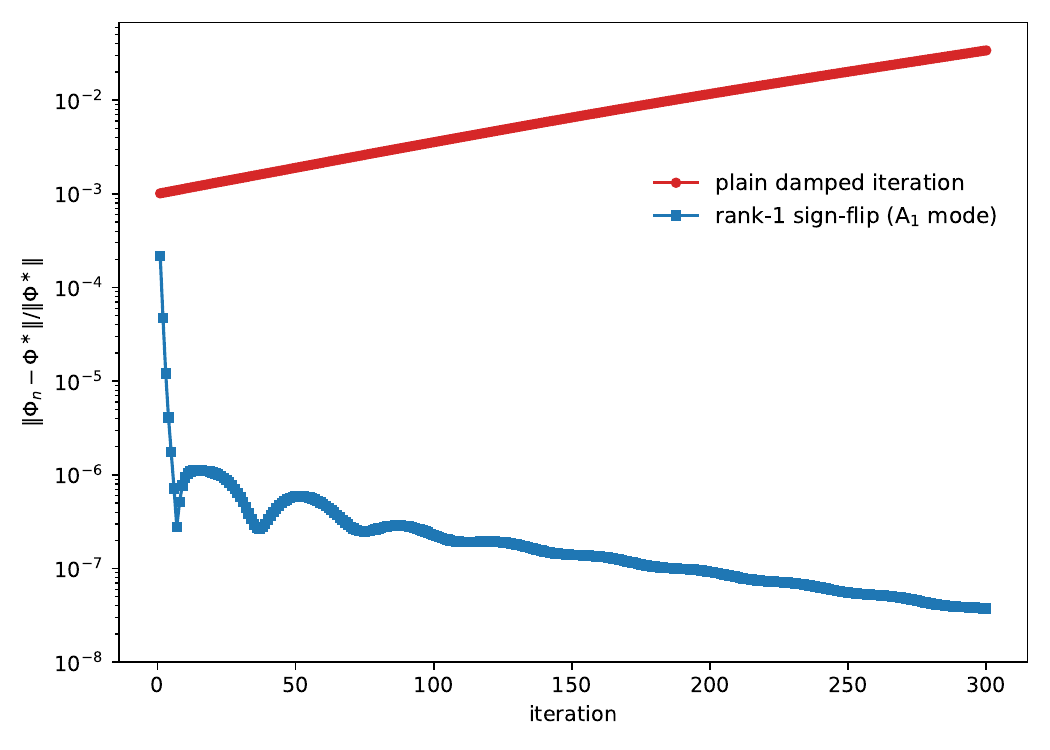}
  \caption{\label{fig:signflip}
  Rank-one sign-flip at the parquet wall ($L=8, \beta=0.65$ for a PS sweep based on the exactly known Monte Carlo vertex):
  distance to the physical fixed point $\Phi^\ast$ versus iteration, seeded along the single unstable
  $A_1$ direction. Plain damped iteration diverges; reversing the damping sign on that one mode makes it
  converge~\cite{Essl2026}.}
\end{figure}

At its onset the fixed-point iteration instability is a single real mode. At $\beta = 0.65$ the Jacobian of the PS sweep at the physical point $\Phi^\ast$ (which was reached by Newton continuation seeded with the measured solution) has exactly one
eigenvalue crossing the real axis at $+1$ ($\mu=1.066$), and its eigenvector is $99.8\%$ $A_1$ at zero transfer, i.e.\ the energy soft mode of Fig.~\ref{fig:irrep}. Following Ref.~\cite{Essl2026}, reversing
the sign of the damping on this one direction (a rank-one projector) turns the repeller into an attractor: plain damped iteration seeded near $\Phi^\ast$ diverges along the mode, whereas the
sign-flipped map converges (Fig.~\ref{fig:signflip}), validating the mechanism directly on a lattice field theory. Below the wall ($\beta\le0.64$) the fixed point is already an attractor and plain damped iteration suffices; like Newton-Krylov, the sign flip is needed only at $\beta=0.65$, the one coupling where the fixed point has just turned repulsive.

We then tried, unsuccessfully,  to extend this mechanism to larger $\beta$:  the unstable subspace proliferates and
turns complex ($1\to6\to8$ eigenvalues with $\mathrm{Re}\,\mu>1$ at $\beta=0.65,0.66,0.67$), so a single
real sign-flip no longer suffices and reaching the thermodynamic $\beta_c$ would require the fuller multi-mode machinery of
Ref.~\cite{Essl2026}, which we do not pursue. However, by comparing with Fig.~\ref{fig:fss}(c) it may well be that $\beta > 0.64$ is in the finite-size ordered phase,
which is outside the scope of this paper; in other words,  it might  not be meaningful to solve the PS / PSD sweeps beyond $\beta > 0.64$ in which case the mechanism of Ref.~\cite{Essl2026} works fine throughout the disordered phase.

\begin{figure}
  \includegraphics[width=\columnwidth]{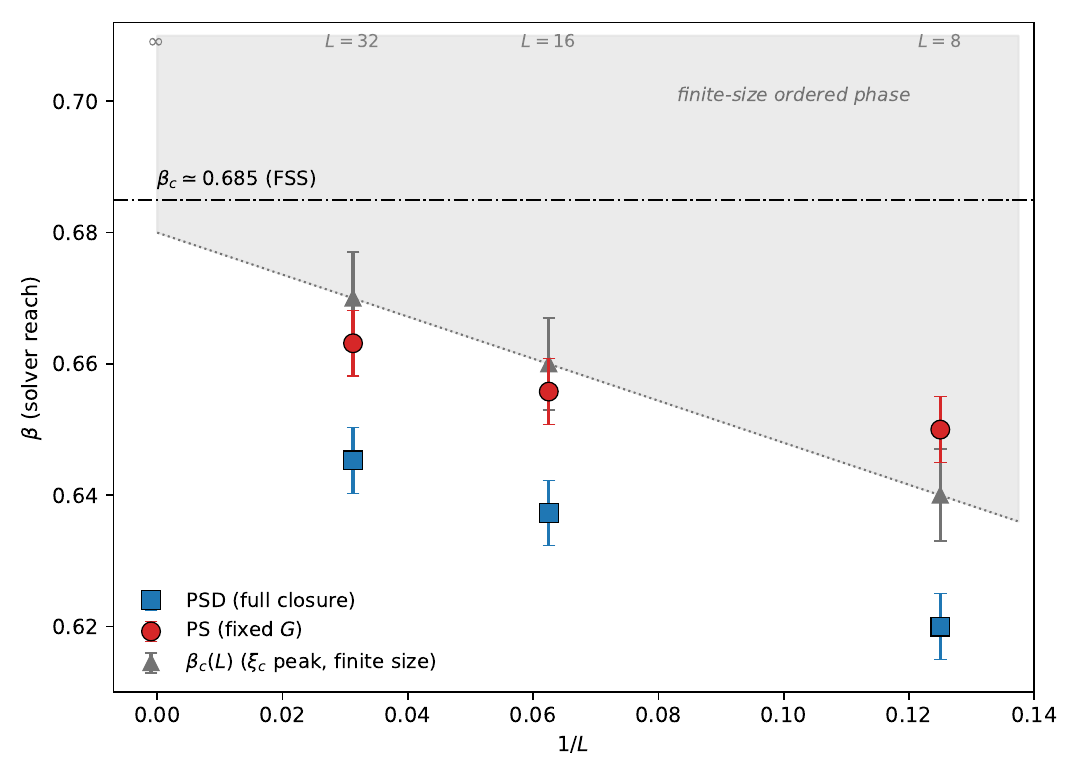}
  \caption{\label{fig:wall}
  Finite-size scaling of the parquet convergence window: $\beta_{\rm wall}$ versus $1/L$ for $L=8,16,32$.
  Blue squares: the wall of the full self-consistent parquet-SDE-Dyson (PSD) map; red circles: the wall of
  the fixed-propagator parquet-SDE (PS) map. Both maps were solved directly only at $L=8$ (PSD wall
  $0.62$; PS wall $0.65$, the single-$A_1$-mode repeller onset that the sign flip reaches); at $L=16,32$
  the wall is located as the coupling at which the $A_1$ $q=0$ ladder eigenvalue reaches its respective
  $L=8$ threshold value (see text). Grey triangles: the finite-size critical point $\beta_c(L)$ from the
  $\xi_c$ peak of Fig.~\ref{fig:fss}(c); the shaded region above is the finite-size ordered phase. Both
  walls drift up with $L$ toward the thermodynamic $\beta_c\simeq0.685$ (dash-dotted). Error bars: the
  $\beta$-grid resolution ($\pm0.005$) for the walls, and the $\xi_c$-peak systematic ($\pm0.007$) for
  $\beta_c(L)$. }
  \end{figure}

There is an argument to support that the location of the convergence window might itself be a finite-size effect.
Because the unstable direction is the $A_1$ zero-transfer mode, the wall (by which we mean the boundary of the convergence window) can be tracked cheaply without solving the PS sweeps (which is very expensive) through the measured $A_1$
ladder eigenvalue $\lambda_{A_1}(q{=}0)$ (which is, we recall, the leading eigenvalue of $\chi_0^{1/2}\Gamma_{\rm ph}\chi_0^{1/2}$
at zero transfer, see Fig.~\ref{fig:irrep}).
The wall is estimated from the coupling at which $\lambda_{A_1}$ reaches the value it takes at the $L=8$ wall. Calibrated
on  PS-sweeps at  $L=8$ and $\beta=0.65$
($\lambda_{A_1, {\rm PS}}^\ast=0.537$), and on the PSD-sweeps at $L=8$ and  $\beta=0.62$ ($\lambda_{A_1, {\rm PSD}}^\ast=0.351$), and assuming that the same $\lambda_{A_1, {\rm PS(D)}}^\ast$ thresholds remain valid at larger system size, we see that  the wall moves \emph{up} monotonically with system size (Fig.~\ref{fig:wall}): $\beta_{\rm wall}=0.650,\,0.656,\,0.663$ (PS) and $0.620,\,0.637,\,0.645$ (PSD) at $L=8,16,32$. Of the two calibrations, the PSD one is cleaner: its $L=8$ wall ($\beta=0.62$) sits safely inside the disordered phase, whereas the PS calibration point ($\beta=0.65$) already lies at or above the finite-size critical coupling $\beta_c(L{=}8)\approx0.64$ (Fig.~\ref{fig:fss}(c)), where the single-mode picture is beginning to break down; the PS drift should be read accordingly.

A linear extrapolation in $1/L$ places
both near $0.65$--$0.67$, still climbing toward the finite-size-scaling estimate
$\beta_c\simeq0.685$. Since the fixed
threshold is a lower bound on the drift (larger $L$ carries more soft-mode weight), this probably
understates how far the \emph{onset} of the window drifts. It would, however, be too strong to conclude that the window opens all the way to $\beta_c$: 
On the contrary, as $\beta\to\beta_c$ the critical channel softens as a whole and the unstable set likely proliferates into many, increasingly complex modes ($1\to6\to8$, above), i.e.\ a manifestation of critical slowing down. 

\bibliography{vertex}

@article{BrowerTamayo1989,
  author  = {Brower, R. C. and Tamayo, P.},
  title   = {Embedded {D}ynamics for $\phi^4$ {T}heory},
  journal = {Phys. Rev. Lett.},
  volume  = {62},
  pages   = {1087},
  year    = {1989},
  doi     = {10.1103/PhysRevLett.62.1087}
}

@article{Salpeter1951,
  author  = {Salpeter, E. E. and Bethe, H. A.},
  title   = {A {R}elativistic {E}quation for {B}ound-{S}tate {P}roblems},
  journal = {Phys. Rev.},
  volume  = {84},
  pages   = {1232},
  year    = {1951},
  doi     = {10.1103/PhysRev.84.1232}
}

@article{BaymKadanoff1961,
  author  = {Baym, G. and Kadanoff, L. P.},
  title   = {Conservation {L}aws and {C}orrelation {F}unctions},
  journal = {Phys. Rev.},
  volume  = {124},
  pages   = {287},
  year    = {1961},
  doi     = {10.1103/PhysRev.124.287}
}

@article{Baym1962,
  author  = {Baym, G.},
  title   = {Self-{C}onsistent {A}pproximations in {M}any-{B}ody {S}ystems},
  journal = {Phys. Rev.},
  volume  = {127},
  pages   = {1391},
  year    = {1962},
  doi     = {10.1103/PhysRev.127.1391}
}

@article{Onsager1944,
  author  = {Onsager, L.},
  title   = {Crystal {S}tatistics. {I}. {A} {T}wo-{D}imensional {M}odel with an {O}rder-{D}isorder {T}ransition},
  journal = {Phys. Rev.},
  volume  = {65},
  pages   = {117},
  year    = {1944},
  doi     = {10.1103/PhysRev.65.117}
}

@article{Guerrero2015,
  author  = {Guerrero, A. I. and Stariolo, D. A. and Almarza, N. G.},
  title   = {Nematic phase in the {$J_1$-$J_2$} square-lattice {I}sing model in an external field},
  journal = {Phys. Rev. E},
  volume  = {91},
  pages   = {052123},
  year    = {2015},
  doi     = {10.1103/PhysRevE.91.052123}
}

@article{Diatlov1957,
  author  = {Diatlov, I. T. and Sudakov, V. V. and Ter-Martirosian, K. A.},
  title   = {Asymptotic meson-meson scattering theory},
  journal = {Sov. Phys. JETP},
  volume  = {5},
  pages   = {631},
  year    = {1957}
}

@article{DeDominicis1964,
  author  = {De Dominicis, C. and Martin, P. C.},
  title   = {Stationary {E}ntropy {P}rinciple and {R}enormalization in {N}ormal and {S}uperfluid {S}ystems. {I.} {A}lgebraic {F}ormulation},
  journal = {J. Math. Phys.},
  volume  = {5},
  pages   = {14},
  year    = {1964},
  doi     = {10.1063/1.1704062}
}

@article{Bickers1991,
  author  = {Bickers, N. E. and White, S. R.},
  title   = {Conserving approximations for strongly fluctuating electron systems. {II.} {N}umerical results and parquet extension},
  journal = {Phys. Rev. B},
  volume  = {43},
  pages   = {8044},
  year    = {1991},
  doi     = {10.1103/PhysRevB.43.8044}
}

@article{BickersScalapino1992,
  author  = {Bickers, N. E. and Scalapino, D. J.},
  title   = {Critical behavior of electronic parquet solutions},
  journal = {Phys. Rev. B},
  volume  = {46},
  pages   = {8050},
  year    = {1992},
  doi     = {10.1103/PhysRevB.46.8050}
}

@article{Toschi2007,
  author  = {Toschi, A. and Katanin, A. A. and Held, K.},
  title   = {Dynamical vertex approximation: A step beyond dynamical mean-field theory},
  journal = {Phys. Rev. B},
  volume  = {75},
  pages   = {045118},
  year    = {2007},
  doi     = {10.1103/PhysRevB.75.045118}
}

@article{Rohringer2018,
  author  = {Rohringer, G. and Hafermann, H. and Toschi, A. and Katanin, A. A. and Antipov, A. E. and Katsnelson, M. I. and Lichtenstein, A. I. and Rubtsov, A. N. and Held, K.},
  title   = {Diagrammatic routes to nonlocal correlations beyond dynamical mean field theory},
  journal = {Rev. Mod. Phys.},
  volume  = {90},
  pages   = {025003},
  year    = {2018},
  doi     = {10.1103/RevModPhys.90.025003}
}

@article{Kozik2015,
  author  = {Kozik, E. and Ferrero, M. and Georges, A.},
  title   = {Nonexistence of the {L}uttinger-{W}ard {F}unctional and {M}isleading {C}onvergence of {S}keleton {D}iagrammatic {S}eries for {H}ubbard-{L}ike {M}odels},
  journal = {Phys. Rev. Lett.},
  volume  = {114},
  pages   = {156402},
  year    = {2015},
  doi     = {10.1103/PhysRevLett.114.156402}
}

@article{Schaefer2016,
  author  = {Sch\"afer, T. and Ciuchi, S. and Wallerberger, M. and Thunstr\"om, P. and Gunnarsson, O. and Sangiovanni, G. and Rohringer, G. and Toschi, A.},
  title   = {Nonperturbative landscape of the {M}ott-{H}ubbard transition: Multiple divergence lines around the critical endpoint},
  journal = {Phys. Rev. B},
  volume  = {94},
  pages   = {235108},
  year    = {2016},
  doi     = {10.1103/PhysRevB.94.235108}
}

@article{Gunnarsson2017,
  title = {Breakdown of {T}raditional {M}any-{B}ody {T}heories for {C}orrelated {E}lectrons},
  author = {Gunnarsson, O. and Rohringer, G. and Sch\"afer, T. and Sangiovanni, G. and Toschi, A.},
  journal = {Phys. Rev. Lett.},
  volume = {119},
  issue = {5},
  pages = {056402},
  numpages = {5},
  year = {2017},
  month = {Aug},
  publisher = {American Physical Society},
  doi = {10.1103/PhysRevLett.119.056402},
  url = {https://link.aps.org/doi/10.1103/PhysRevLett.119.056402}
}

@article{Gunnarsson2016,
  title = {Parquet decomposition calculations of the electronic self-energy},
  author = {Gunnarsson, O. and Sch\"afer, T. and LeBlanc, J. P. F. and Merino, J. and Sangiovanni, G. and Rohringer, G. and Toschi, A.},
  journal = {Phys. Rev. B},
  volume = {93},
  issue = {24},
  pages = {245102},
  numpages = {17},
  year = {2016},
  month = {Jun},
  publisher = {American Physical Society},
  doi = {10.1103/PhysRevB.93.245102},
  url = {https://link.aps.org/doi/10.1103/PhysRevB.93.245102}
}

@article{LinLindsey2018,
  author  = {Lin, Lin and Lindsey, Michael},
  title   = {Variational structure of {L}uttinger-{W}ard formalism and bold diagrammatic expansion for {E}uclidean lattice field theory},
  journal = {Proc. Natl. Acad. Sci. U.S.A.},
  volume  = {115},
  pages   = {2282},
  year    = {2018},
  doi     = {10.1073/pnas.1720782115}
}

@article{Eckhardt2023,
	title = {A functional-analysis derivation of the parquet equation},
	pages = {203},
	author = {Eckhardt, Christian J. and Kappl, Patrick and Kauch, Anna and Held, Karsten},
	journal = {SciPost Phys.},
	volume = {15},
	year = {2023},
	publisher = {SciPost},
	doi = {10.21468/SciPostPhys.15.5.203},
	url = {https://scipost.org/10.21468/SciPostPhys.15.5.203}
}

@misc{Essl2026,
      title={Stabilizing the parquet problem}, 
      author={Herbert Eßl and Stefan Rohshap and Marcel Gievers and Markus Wallerberger and Alessandro Toschi and Anna Kauch},
      year={2026},
      eprint={2606.04936},
      archivePrefix={arXiv},
      primaryClass={cond-mat.str-el},
      url={https://arxiv.org/abs/2606.04936}, 
}

@article{ALPSCore1,
        title = {Updated core libraries of the {ALPS} project},
        volume = {213},
        issn = {0010-4655},
        url = {https://www.sciencedirect.com/science/article/pii/S0010465516303885},
        doi = {10.1016/j.cpc.2016.12.009},
        urldate = {2022-09-21},
        journal = {Computer Physics Communications},
        author = {Gaenko, A. and Antipov, A. E. and Carcassi, G. and Chen, T. and Chen, X. and Dong, Q. and Gamper, L. and Gukelberger, J. and Igarashi, R. and Iskakov, S. and Könz, M. and LeBlanc, J. P. F. and Levy, R. and Ma, P. N. and Paki, J. E. and Shinaoka, H. and Todo, S. and Troyer, M. and Gull, E.},
        month = apr,
        year = {2017},
        pages = {235--251},
}

@techreport{ALPSCore2,
        title = {Updated {Core} {Libraries} of the {ALPS} {Project}},
        url = {http://arxiv.org/abs/1811.08331},
        number = {arXiv:1811.08331},
        urldate = {2022-09-21},
        institution = {arXiv},
        author = {Wallerberger, Markus and Iskakov, Sergei and Gaenko, Alexander and Kleinhenz, Joseph and Krivenko, Igor and Levy, Ryan and Li, Jia and Shinaoka, Hiroshi and Todo, Synge and Chen, Tianran and Chen, Xi and LeBlanc, James P. F. and Paki, Joseph E. and Terletska, Hanna and Troyer, Matthias and Gull, Emanuel},
        month = nov,
        year = {2018},
        doi = {10.48550/arXiv.1811.08331},
}

@article{Cooper1982,
title = {Solving $\phi_{1,2}^4$ field theory with {M}onte {C}arlo},
journal = {Nuclear Physics B},
volume = {210},
number = {2},
pages = {210-228},
year = {1982},
issn = {0550-3213},
doi = {https://doi.org/10.1016/0550-3213(82)90240-1},
url = {https://www.sciencedirect.com/science/article/pii/0550321382902401},
author = {Fred Cooper and B. Freedman and Dean Preston}
}

@article{RMP_fRG,
  title = {Functional renormalization group approach to correlated fermion systems},
  author = {Metzner, Walter and Salmhofer, Manfred and Honerkamp, Carsten and Meden, Volker and Sch\"onhammer, Kurt},
  journal = {Rev. Mod. Phys.},
  volume = {84},
  issue = {1},
  pages = {299--352},
  numpages = {0},
  year = {2012},
  month = {Mar},
  publisher = {American Physical Society},
  doi = {10.1103/RevModPhys.84.299},
  url = {https://link.aps.org/doi/10.1103/RevModPhys.84.299}
}

@article{Berges2002,
title = {Non-perturbative renormalization flow in quantum field theory and statistical physics},
journal = {Physics Reports},
volume = {363},
number = {4},
pages = {223-386},
year = {2002},
note = {Renormalization group theory in the new millennium. IV},
issn = {0370-1573},
doi = {https://doi.org/10.1016/S0370-1573(01)00098-9},
url = {https://www.sciencedirect.com/science/article/pii/S0370157301000989},
author = {Jürgen Berges and Nikolaos Tetradis and Christof Wetterich}
}

@article{SDE,
title = {Dyson-{S}chwinger equations and their application to hadronic physics},
journal = {Progress in Particle and Nuclear Physics},
volume = {33},
pages = {477-575},
year = {1994},
issn = {0146-6410},
doi = {https://doi.org/10.1016/0146-6410(94)90049-3},
url = {https://www.sciencedirect.com/science/article/pii/0146641094900493},
author = {Craig D. Roberts and Anthony G. Williams}
}

@article{Knoll2004,
title = {Jacobian-free {N}ewton–{K}rylov methods: a survey of approaches and applications},
journal = {Journal of Computational Physics},
volume = {193},
number = {2},
pages = {357-397},
year = {2004},
issn = {0021-9991},
doi = {https://doi.org/10.1016/j.jcp.2003.08.010},
url = {https://www.sciencedirect.com/science/article/pii/S0021999103004340},
author = {D.A. Knoll and D.E. Keyes}
}

@article{WalkerHomerNi,
author = {Walker, Homer F. and Ni, Peng},
title = {Anderson {A}cceleration for {F}ixed-{P}oint {I}terations},
journal = {SIAM Journal on Numerical Analysis},
volume = {49},
number = {4},
pages = {1715-1735},
year = {2011},
doi = {10.1137/10078356X},
URL = {https://doi.org/10.1137/10078356X},
eprint = {https://doi.org/10.1137/10078356X}
}

@article{FangSaad2009,
author = {Fang, Haw-ren and Saad, Yousef},
title = {Two classes of multisecant methods for nonlinear acceleration},
journal = {Numerical Linear Algebra with Applications},
volume = {16},
number = {3},
pages = {197-221},
doi = {https://doi.org/10.1002/nla.617},
url = {https://onlinelibrary.wiley.com/doi/abs/10.1002/nla.617},
eprint = {https://onlinelibrary.wiley.com/doi/pdf/10.1002/nla.617},
year = {2009}
}

@article{Berges2004,
  title = {$n$-particle irreducible effective action techniques for gauge theories},
  author = {Berges, J\"urgen},
  journal = {Phys. Rev. D},
  volume = {70},
  issue = {10},
  pages = {105010},
  numpages = {21},
  year = {2004},
  month = {Nov},
  publisher = {American Physical Society},
  doi = {10.1103/PhysRevD.70.105010},
  url = {https://link.aps.org/doi/10.1103/PhysRevD.70.105010}
}

@Article{Carrington2004,
author={Carrington, M. E.},
title={The {4PI} effective action for $\phi^4$theory},
journal={The European Physical Journal C - Particles and Fields},
year={2004},
month={Jun},
day={01},
volume={35},
number={3},
pages={383-392},
issn={1434-6052},
doi={10.1140/epjc/s2004-01849-6},
url={https://doi.org/10.1140/epjc/s2004-01849-6}
}

@book{AGD,
  author    = {Abrikosov, A. A. and Gorkov, L. P. and Dzyaloshinskii, I. E.},
  title     = {Methods of Quantum Field Theory in Statistical Physics},
  publisher = {Dover Publications},
  address   = {Mineola, NY},
  year      = {1963}
}

@book{FetterWalecka,
  author    = {Fetter, Alexander L. and Walecka, John D.},
  title     = {Quantum Theory of Many-Particle Systems},
  publisher = {McGraw--Hill},
  address   = {New York},
  year      = {1971}
}

@book{NegeleOrland,
  author    = {Negele, John W. and Orland, Henri},
  title     = {Quantum Many-Particle Systems},
  publisher = {Addison--Wesley},
  address   = {Redwood City, CA},
  year      = {1988}
}

\end{document}